\documentclass[11pt]{article}
\usepackage[utf8]{inputenc}
\usepackage{amsfonts}
\usepackage{amsmath}
\usepackage{amssymb}
\usepackage{indentfirst}
\usepackage{graphicx}
\usepackage[dvipsnames]{xcolor}
\usepackage[colorlinks]{hyperref}
\usepackage{cite}
\usepackage{array}
\usepackage{microtype}
\usepackage{soul}
\usepackage[normalem]{ulem}
\usepackage{cancel}
\usepackage{soul}
\usepackage{xspace}
\usepackage{colortbl}
\usepackage{caption}
\newcolumntype{C}[1]{>{\centering\arraybackslash}p{#1}}
\definecolor{maroon}{cmyk}{0,0.87,0.68,0.32}

\numberwithin{equation}{section}
\allowdisplaybreaks

\makeatletter

\begin{document}

\begin{titlepage}
\vspace{3cm}

\baselineskip=24pt

\begin{center}
\textbf{\LARGE Maxwell kinematical algebras and 3D gravities}
\par\end{center}{\LARGE \par}

\begin{center}
	\vspace{1cm}
	\textbf{Patrick Concha}$^{\ast, \bullet}$,
        \textbf{Nelson Gallegos}$^{\star, \bullet}$
        \textbf{Evelyn Rodríguez}$^{\ast, \bullet}$
        \textbf{Sebastián Salgado}$^{\dag}$
	\small
	\\[5mm]
    $^{\ast}$\textit{Departamento de Matemática y Física Aplicadas, }\\
	\textit{ Universidad Católica de la Santísima Concepción, }\\
\textit{ Alonso de Ribera 2850, Concepción, Chile.}
\\[2mm]
$^{\bullet}$\textit{Grupo de Investigación en Física Teórica, GIFT, }\\
	\textit{ Universidad Católica de la Santísima Concepción, }\\
\textit{ Alonso de Ribera 2850, Concepción, Chile.}
\\[2mm]
$^{\star}$\textit{Departamento de Física, Universidad de Concepción,}\\
	\textit{Casilla 160-C, Concepción, Chile.}
\\[2mm]
	$^{\dag}$\textit{Dirección de Investigación, Universidad Bernardo O'Higgins, }\\
	\textit{General Gana 1702, Santiago, Chile.}
	\\[5mm]
	\footnotesize
	\texttt{patrick.concha@ucsc.cl},
        \texttt{ngallegos2020@udec.cl}
        \texttt{erodriguez@ucsc.cl},
        \texttt{sebasalg@gmail.com},
	\par\end{center}
\vskip 26pt
\begin{abstract}

In this paper, we present a Maxwell extension of kinematical Lie algebras by promoting the contraction method underlying the Bacry and Lévy-Leblond cube to a semigroup expansion framework. Within this approach, we show that both non- and ultra-relativistic Maxwell algebras admitting non-degenerate invariant bilinear forms can be systematically obtained from different parent algebras through a unified expansion scheme, leading to a Maxwellian kinematical cube. This construction is further generalized to an infinite hierarchy of kinematical algebras. The expansion method naturally provides the corresponding invariant tensors, allowing for the systematic construction of three-dimensional Chern–Simons gravity theories.

\end{abstract}
\end{titlepage}\newpage {} 

{\baselineskip=12pt \tableofcontents{}}

\section{Introduction}
Kinematical Lie algebras encode the transformations between inertial observers in spacetime under the assumptions of homogeneity, isotropy, invariance under parity and time reversal, and the requirement that the transformations relating different inertial frames form
non-compact subgroups. These algebras were classified by Bacry and Lévy-Leblond, who showed that they can be organized as the vertices of a cube \cite{Bacry:1968zf}.

Kinematical algebras naturally split into relativistic and non-Lorentzian symmetry structures. In recent years, non-Lorentzian regimes of (super)gravity theories have attracted increasing attention due to their relevance in a wide range of physical contexts. On the one hand, non-relativistic (Galilean) symmetries play a central role in holography \cite{Son:2008ye,Balasubramanian:2008dm,Kachru:2008yh,Taylor:2008tg,Bagchi:2009my,Hartnoll:2009sz,Bagchi:2009pe,Christensen:2013lma,Christensen:2013rfa,Hartong:2014oma,Hartong:2014pma,Zaanen:2015oix}, Ho\v{r}ava-Lifshitz gravity \cite{Horava:2009uw,Hartong:2015wxa,Hartong:2015zia,Taylor:2015glc,Hartong:2016yrf,Devecioglu:2018apj} and effective descriptions of condensed matter systems such as the quantum Hall effect \cite{Hoyos:2011ez,Son:2013rqa,Abanov:2014ula,Geracie:2014nka,Gromov:2015fda}. On the other hand, ultra-relativistic (Carrollian) symmetries have emerged in diverse settings, ranging from tensionless strings and warped CFTs \cite{Hofman:2014loa,Bagchi:2013bga,Bagchi:2018wsn} to asymptotic symmetries, flat holography, and near-horizon physics \cite{Duval:2014uva,Hartong:2015xda,Hartong:2015usd,Bagchi:2016bcd,Donnay:2019jiz,Ciambelli:2019lap,Grumiller:2019fmp,Perez:2021abf,Donnay:2022aba,Perez:2022jpr,Fuentealba:2022gdx,Saha:2023hsl,Nguyen:2023vfz,deBoer:2023fnj,Ecker:2023uwm,Donnay:2023mrd,Bagchi:2023cen}. 

A  good laboratory to probe structural aspects of gravity in the non-Lorentzian regime is provided by the three-dimensional model formulated in the Chern-Simons (CS) framework, which share several features with higher-dimensional gravity theories, such as black-hole solutions and their associated thermodynamics \cite{Banados:1992wn}. In this context, the existence of a non-degenerate invariant bilinear form is a crucial requirement, as it guarantees well-defined field equations. This condition imposes non-trivial constraints on the underlying symmetry algebra. For instance, in the non-relativistic case, a consistent CS action requires the extended Bargmann algebra \cite{Papageorgiou:2009zc,Bergshoeff:2016lwr}, which corresponds to a double central extension of the Galilei algebra. In contrast, in the ultra-relativistic regime, the Carroll algebra already admits a non-degenerate invariant tensor \cite{Matulich:2019cdo,Ravera:2019ize,Ali:2019jjp,Concha:2023bly,Concha:2024tcu}. More recently, a non-vanishing torsion has been incorporated into three-dimensional Carrollian
gravity, affecting the non-affinity of null generators and boundary dynamics \cite{Concha:2025vhd}. Matter couplings and conformal approaches to Carroll gravity have also been explored in \cite{Bergshoeff:2017btm,Hansen:2021fxi,Baiguera:2022lsw,Bergshoeff:2023vfd,Bergshoeff:2024ilz,Bergshoeff:2026cxt,Bekaert:2026cvx}.

At the relativistic level, the Maxwell algebra has attracted considerable interest over the past decades, which has been introduced to describe a constant electromagnetic field in a Minkowski background \cite{Bacry:1970ye,Bacry:1970du,Schrader:1972zd,Gomis:2017cmt}. The Maxwell algebra can be understood as an extension and deformation of the Poincaré algebra and has been extensively studied within the three-dimensional CS gravity framework \cite{Salgado:2014jka,Hoseinzadeh:2014bla,Concha:2018zeb,Concha:2023nou,Aviles:2025nah}. In particular, recent analyses of asymptotically flat cosmological solutions in
Maxwell CS gravity have shown that the gravitational Maxwell field modifies the structure of canonical generators and contributes non-trivially to the first law of thermodynamics \cite{Aviles:2025nah}. Supersymmetric extensions of the Maxwell algebra, as well as higher-spin generalizations, have also been explored \cite{deAzcarraga:2014jpa,Concha:2014tca,Penafiel:2017wfr,Ravera:2018vra,Concha:2018jxx,Caroca:2021bjo,Matulich:2023xpw,Concha:2024rac}. In contrast to the Poincaré case, the non-Lorentzian counterpart of the Maxwell CS gravity requires a more subtle treatment, as both the non- and ultra-relativistic regimes generically lead to degeneracies in the invariant bilinear form. To overcome this issue, a Maxwell extension of the extended Bargmann algebra has been introduced in \cite{Aviles:2018jzw}, obtained as a contraction of the [Maxwell]$\oplus \mathfrak{u}(1)^{3}$ algebra. Similarly, in the ultra-relativistic regime, a Maxwellian extended Carroll algebra has been proposed in order to ensure a non-degenerate formulation \cite{Concha:2021jnn}.

In this work, we show that the non-degenerate non-Lorentzian Maxwell algebras introduced in \cite{Aviles:2018jzw,Concha:2021jnn} can be naturally understood as elements of a Maxwellian
generalization of the Bacry and L\'{e}vy-Leblond cube, obtained by replacing the
In\"{o}n\"{u}-Wigner contraction with an expansion procedure. In contrast to contractions, expansion methods \cite{Hatsuda:2001pp,deAzcarraga:2002xi,Izaurieta:2006zz,deAzcarraga:2007et} typically increase the number of generators of the original algebra. In particular, the semigroup expansion method \cite{Izaurieta:2006zz} has proven to be a powerful tool for constructing non-Lorentzian symmetry algebras that admit a non-degenerate invariant bilinear form \cite{Concha:2023bly}, and therefore for defining consistent non-Lorentzian (super)gravity theories \cite{deAzcarraga:2019mdn,Bergshoeff:2019ctr,Concha:2019lhn,Bergshoeff:2020fiz,Concha:2020eam,Concha:2022muu,Concha:2022jdc,Concha:2023ejs,Concha:2024dap}. Remarkably, within this framework it is also possible to preserve the dimensionality of the original algebra. A particularly illustrative example is provided by the semigroup $S_{E}^{(1)}$, which reproduces the Inönü–Wigner contraction for an appropriate choice of subspace decomposition. This observation allows one to reinterpret the Bacry and Lévy–Leblond cube~\ref{fig1a} as a diagram in which the arrows correspond to $S_{E}^{(1)}$ expansions rather than contractions \cite{Concha:2023bly}. It naturally motivates the construction of a generalized cube, where each arrow is replaced by an expansion associated with a higher-order semigroup.

The paper is organized as follows. In section~\ref{sec2} we briefly review the kinematical algebras introduced in \cite{Bacry:1968zf}. Our main results are presented in sections~\ref{sec3} and~\ref{sec4}. In section~\ref{sec3} we construct the Maxwellian generalization of the kinematical algebras using the semigroup expansion procedure and analyze the corresponding CS gravity actions. In section~\ref{sec4} we extend this construction to a broader class of generalized kinematical algebras based on an arbitrary semigroup $S_{E}^{(N)}$. Finnaly, section~\ref{sec5} contains our conclusions and a discussion of possible future directions.

\section{Kinematical Lie algebras}\label{sec2}

In this section we briefly review the kinematical Lie algebras classified by Bacry and Lévy-Leblond in \cite{Bacry:1968zf}. By construction, kinematical algebras do not
describe the dynamics of particles or their interactions; rather, they specify
how reference frames move relative to each other in different spacetimes.
Nevertheless, spacetime symmetry algebras play a central role in gravitational theories, where they are often promoted to gauge symmetries. A notable example is three-dimensional gravity, which admits a formulation as a Chern-Simons (CS) gauge theory for the AdS group \cite{Achucarro:1987vz,Witten:1988hc,Zanelli:2005sa}. 

Motivated by this perspective, a systematic study of gauge
theories associated with the non- and ultra-relativistic limits of relativistic symmetry algebras becomes particularly relevant for understanding non-Lorentzian regimes of gravity. In three spacetime dimensions, such theories are naturally described within the CS framework, where the fundamental field is a one-form
gauge connection $A$ valued in a Lie algebra $\mathfrak{g}$. The action is obtained by integrating the standard CS form,
\begin{equation}
S_{\mathrm{CS}}=\frac{k}{4\pi}\int\left\langle A\mathrm{d}A+\frac{2}{3}%
A^{3}\right\rangle\,,\label{csgeneral}%
\end{equation}
where $\left\langle \cdots\right\rangle $ denotes an invariant, symmetric bilinear form on $\mathfrak{g}$. The non-degeneracy of this invariant tensor ensures that the equations of motion enforce the vanishing of the gauge curvature $F=\mathrm{d}A+A^2$. As a consequence, the existence of a non-degenerate invariant bilinear form becomes a crucial ingredient in the construction of CS gravity theories with well-defined field equations, and must therefore be analyzed on a case-by-case basis for each algebra under consideration \cite{Matulich:2019cdo,Concha:2023bly}.

Let us consider the three-dimensional AdS algebra $\mathfrak{so}(2,2)$, which serves as the starting point for the construction of the Bacry and Lévy-Leblond cube. This
algebra is spanned by the generators $\{J_{A},P_{A}\}$, where the
indices $A,B=0,1,2$ are raised and lowered with the Minkowski metric
$\eta_{AB}=\mathrm{diag}(-,+,+)$. These generators satisfy the commutation relations:
\begin{align}
\left[  \tilde{J}_{A},\tilde{J}_{B}\right]  &=\epsilon_{ABC}\tilde{J}%
^{C}\,, & \left[  \tilde{J}_{A},\tilde{P}_{B}\right]
&=\epsilon_{ABC}\tilde{P}^{C}\,, & \left[  \tilde{P}_{A},\tilde
{P}_{B}\right]  &=\epsilon_{ABC}\tilde{J}^{C}\,,\label{ads}%
\end{align}
where the Lorentzian Levi-Civita pseudotensor is defined by $\epsilon_{012}=1$. The Poincaré algebra is obtained from (\ref{ads}) by
means of the In\"{o}n\"{u}-Wigner contraction, implemented through the rescaling $\tilde{P}_{A}\rightarrow
\ell\tilde{P}_{A}$ followed by the limit $\ell\rightarrow\infty$. This procedure corresponds to the vanishing cosmological constant limit, where the
contraction parameter is related to the cosmological constant as $\ell
^{-2}=-\Lambda$. In this flat limit, the translational sector
becomes abelian, namely \ $\left[  \tilde{P}_{A},\tilde{P}_{B}\right]  =0$. Both the AdS and Poincaré algebras are Lorentzian kinematical algebras, describing transformations between relativistic observers in spacetimes of constant negative and vanishing curvature, respectively.

Non-Lorentzian kinematical algebras arise from the AdS and Poincaré algebras by considering the physical limits in which the speed of light is taken to zero or to infinity. Implementing these non-Lorentzian limits requires splitting the Lorentz-covariant index into temporal and spatial components, $A\rightarrow(0,a)$. Accordingly, the generators are decomposed as
\begin{align}
\tilde{J}_{A}&=\left(  \tilde{J}_{0},\tilde{J}_{a}\right)  \equiv\left(  
J,G_{a}\right) \,, & \tilde{P}_{A}&=\left(  \tilde{P}%
_{0},\tilde{P}_{a}\right)  \equiv\left(  H,P_{a}\right)\,,
\end{align}
where the spatial index takes the values $a=1,2$. Here, $J$ generates spatial rotations in two-dimensional Euclidean space, while $G_{a}$ generates boost
transformations relating space and time for different observers. The generator $H$ corresponds to time translations, and $P_{a}$ generates spatial
translations. In terms of these generators, the AdS and Poincaré algebras can be rewritten as follows:
\begin{align}
\textbf{AdS}: &&\left[  {J},{G}_{a}\right]  &=\epsilon_{ab}{G}^{b}\,,& \left[  {H},{G}_{a}\right]  &=\epsilon_{ab}{P}^{b}\,,&\left[{G}_{a},{G}_{b}\right]&=-\epsilon_{ab}{J}\,,\notag\\
&&\left[  {H},{P}_{a}\right] &=\epsilon_{ab}{G}_{b}\,,&\left[  {J},{P}_{a}\right]  &=\epsilon_{ab}{P}^{b}\,, & \left[  {P}_{a},{P}_{b}\right]&=-\epsilon_{ab}{J}\,,\notag\\
&&\left[ {G}_{a},{P}_{b}\right] & =-\epsilon_{ab}{H}\,,\label{aAdS}\\
\notag\\
\textbf{Poincaré}:&&  \left[ {J},{G}_{a}\right]  &=\epsilon_{ab}{G}^{b}\,, & \left[  {H},{G}_{a}\right]&=\epsilon_{ab}{P}^{b}\,,&\left[  {G}_{a},{G}_{b}\right] & =-\epsilon_{ab}{J}\,,\notag\\
 &&\left[{J},{P}_{a}\right]&=\epsilon_{ab}{P}^{b}\,, & \left[ {G}_{a},{P}_{b}\right]&=-\epsilon_{ab}{H}\,,
\end{align}
where $\epsilon_{ab}\equiv \epsilon_{0ab}$, $\epsilon^{ab}\equiv \epsilon^{0ab}$. The non- and ultra-relativistic limits are introduced through speed–space and speed–time contractions, respectively. These limits are defined by the following rescalings:
\begin{align}
\text{Speed-space rescaling:} &&{P}_{a}&\rightarrow\sigma\,P_{a}\,,& G_{a}&\rightarrow\sigma\,G_{a}\,,\notag\\
 \text{Speed-time rescaling:} &&{G}_{a}&\rightarrow\kappa\,G_{a} \,,&{H}&\rightarrow\kappa\,H\,.
\end{align}
The scaling parameters are related to the speed of light as
$\sigma=c$ and $\kappa=c^{-1}$. Applying both contractions to the AdS and
Poincar\'{e} algebras gives rise to four distinct kinematical Lie algebras. The
non-relativistic versions of these algebras are obtained by considering the
speed-space rescaling in the limit $\sigma\rightarrow\infty$. They are given
by \cite{Bacry:1968zf}:
\begin{align}
    \textbf{Newton-Hooke}:&&\left[  J,G_{a}\right]  &=\epsilon_{ab}G_{b}\,,&\left[  H,G_{a}\right]&=\epsilon_{ab}P_{b}\,,\notag\\
   && \left[J,P_{a}\right] &=\epsilon_{ab}P_{b}\,, &\left[  H,P_{a}\right]&=\epsilon_{ab}G_{b}\,,\\
\notag\\
 \textbf{Galilei}:&& \left[  J,G_{a}\right]  &=\epsilon_{ab}G_{b}\,,&\left[  H,G_{a}\right]&=\epsilon_{ab}P_{b}\,,\notag\\  
 &&\left[  J,P_{a}\right]  &=\epsilon_{ab}P_{b}\,. \label{Gal}
\end{align}
On the other hand, the Carrollian limits of the AdS and Poincar\'{e} algebras are obtained by applying the speed-time rescaling and taking the limit $\kappa\rightarrow\infty$. The resulting  algebras are given by \cite{Bacry:1968zf}:
 \begin{align}
    \textbf{Para-Poincaré}:&& \left[  J,G_{b}\right]  &=\epsilon_{ab}G^{b}\,,&\left[  G_a,P_{a}\right]&=-\epsilon_{ab}H\,,\notag\\
   && \left[  J,P_{a}\right]  &=\epsilon_{ab}P^{b}\,, & \left[  P_{a},P_{b}\right]&=-\epsilon_{ab}J\,, \notag \\
&&\left[  H,P_{a}\right] & =\epsilon_{ab}G_{b}\,,\label{adscar}\\
\notag\\
    \textbf{Carroll}:&&\left[  J,G_{b}\right]  &=\epsilon_{ab}G^{b}\,,&\left[ G_{a},P_{a}\right]&=-\epsilon_{ab}H\,,\notag\\
    &&\left[  J,P_{a}\right]  &=\epsilon_{ab}P^{b}\,.
\end{align}
Moreover, the Para-Poincaré (also known as the AdS-Carroll) and Carroll algebras are related through the vanishing cosmological constant limit. The cosmological constant can be reintroduced by means of the rescaling $(H,P_{a})\rightarrow(\ell H,\ell
P_{a})$ in \eqref{adscar}, followed by the limit $\ell\rightarrow\infty$, which leads to the Carroll algebra. An analogous reasoning can be applied to the Newton-Hooke and Para-Galilei algebras, leading to the Galilei and static algebras, respectively.

\begin{figure}[h!]
\centering
\includegraphics[width=.461\columnwidth]{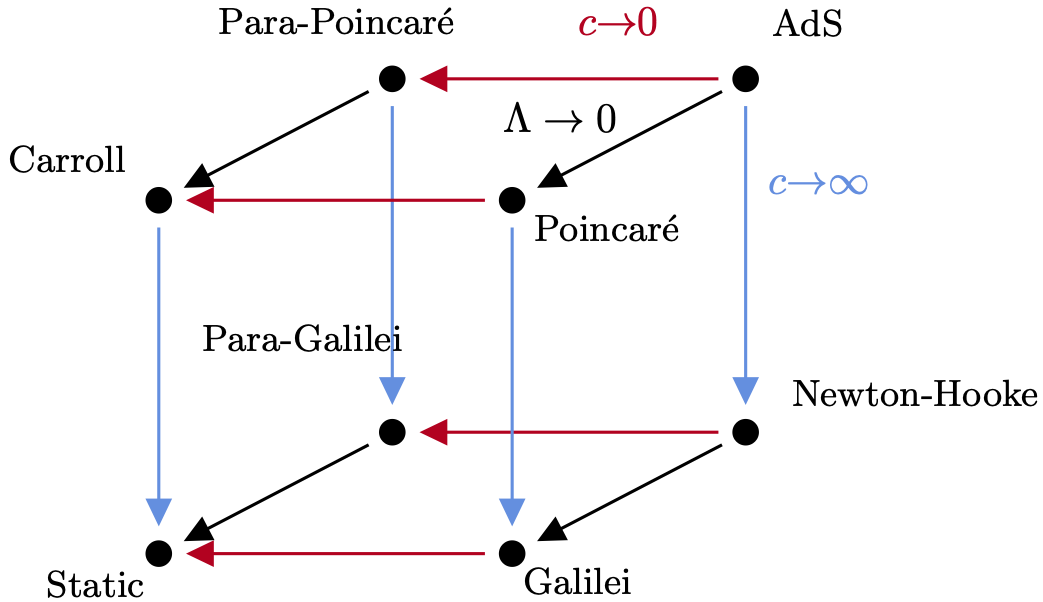}
\captionsetup{font=footnotesize}
\caption{Bacry and Lévy-Leblond cube of kinematical algebras \cite{Bacry:1968zf}.}
\label{fig1a}
\end{figure}

The complete set of contractions is schematically illustrated by the
Bacry and Lévy-Leblond cube \cite{Bacry:1968zf} of kinematical algebras (see Fig.~\ref{fig1a}). Let us note that the cube also includes the so-called static and para-Galilean algebras, which can be obtained as speed-space contractions of the Carroll and AdS–Carroll algebras (or, equivalently, as speed-time contractions of the Galilean and Newton-Hooke algebras). 
\section{Maxwell generalization of kinematical Lie algebras and gravity actions}\label{sec3}
The non-Lorentzian limits of the Maxwell algebra have already been investigated in \cite{Aviles:2018jzw,Concha:2021jnn}, where it was shown that additional central charges must be introduced in both the non- and ultra-relativistic regimes in order to avoid degeneracies in the invariant bilinear form. In this section, we show that both non-degenerate algebras can be embedded into a Maxwellian version of the Bacry and Lévy-Leblond cube \cite{Bacry:1968zf}, in which the standard contraction scheme is replaced by a more general algebraic construction.

With the aim of constructing a Maxwellian version of the kinematical Lie algebras \cite{Bacry:1968zf} without degeneracy, we begin by considering the general structure of an $S_{E}^{(2)}$-expansion, which replaces the usual contraction limit. We introduce a general Lie algebra $\mathfrak{g}=\mathrm{span}\{\mathbf{T}_{A}\}=V_{0}\oplus V_{1}$ endowed with a subspace decomposition
\begin{align}
\lbrack V_{0},V_{0}]&\subset V_{0}\,, &[V_{0},V_{1}]&\subset
V_{1}\,, &[V_{1},V_{1}] &\subset V_{0}\,.\label{decomp}%
\end{align}
We further consider the abelian semigroup
\begin{equation}
    S_{E}^{\left(  2\right)}=\left\{  \lambda_{0},\lambda_{1},\lambda_{2},\lambda_{3}\right\}\,,
\end{equation} 
endowed with the following multiplication law
\begin{equation}
\lambda _{\alpha }\lambda _{\beta }=\left\{ 
\begin{array}{lcl}
\lambda _{\alpha +\beta }\,\,\,\, & \mathrm{if}\,\,\,\,\alpha +\beta \leq
3\,, &  \\ 
\lambda _{3}\,\,\, & \mathrm{if}\,\,\,\,\alpha +\beta >3\,. & 
\end{array}%
\right.   \label{MLSE2}
\end{equation}
It is straightforward to verify that the decomposition (\ref{decomp}) is in
resonance with the splitting of the semigroup $S_{E}^{(2)}$ into the subsets
\begin{align}
    S_{0}&=\{\lambda
_{0},\lambda_{2},\lambda_{3}\}\,, & S_{1}&=\{\lambda_{1},\lambda_{3}\}\,,
\end{align}
which satisfy
\begin{align}
    S_0\cdot S_0&\subset S_0\,,& S_0\cdot S_1&\subset S_1\,,& S_1\cdot S_1\subset S_0\,. \label{rc}
\end{align}
This resonance condition allows for the construction of the corresponding resonant expanded subalgebra
\begin{equation}
    \mathfrak{G}=\left(  S_{0}\times V_{0}\right)  \oplus\left(  S_{1}\times
V_{1}\right) \,. \label{rexp}
\end{equation}
Moreover, since the semigroup admits a zero element $\lambda_{3}$, satisfying $\lambda_3 \lambda_i=\lambda_3$ for any $\lambda_i\in S_{E}^{(2)}$, one can consistently perform a $0_{S}$-reduction by imposing the condition $\lambda_{3}\times \mathbf{T}_{A}=0$. The resulting algebra will be refer as a resonant $0_{S}$-reduced $S_{E}^{\left(  2\right)  }$-expanded algebra.

\begin{figure}[h!]
\centering
\includegraphics[width=.35\columnwidth]{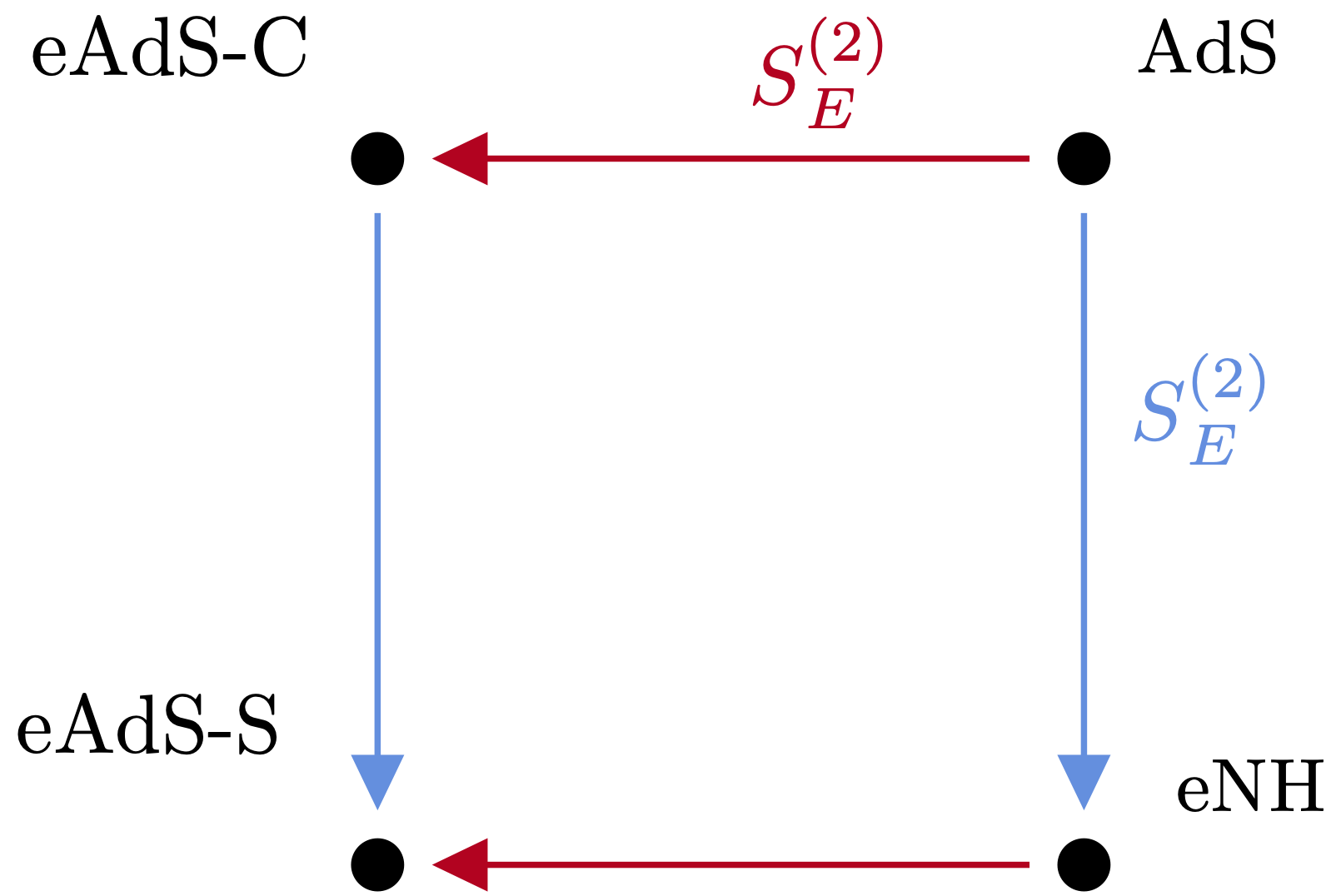}
\captionsetup{font=footnotesize}
\caption{Extended kinematical algebras starting from the AdS algebra \cite{Concha:2023bly}.}
\label{fig2}
\end{figure}

Our starting point is the AdS algebra and its non-Lorentzian counterparts. As shown in \cite{Concha:2023bly, Concha:2024dap}, the $S_{E}^{(2)}$-expansion applied to the AdS algebra, upon considering appropriate speed-space and speed-time subspace decompositions, reproduces a family of extended kinematical algebras (see Fig.~\ref{fig2}). At the non-relativistic level, the resulting algebra corresponds to the extended Newton-Hooke algebra \cite{Aldrovandi:1998im,Gibbons:2003rv,Brugues:2006yd,Alvarez:2007fw,Papageorgiou:2010ud,Duval:2011mi,Duval:2016tzi}, which, in the vanishing cosmological constant limit $\Lambda\rightarrow 0$, reduces to the extended Bargmann algebra \cite{Papageorgiou:2009zc,Bergshoeff:2016lwr}. In the ultra-relativistic regime, the expansion yields to the extended Para-Poincaré algebra (or extended AdS-Carroll algebra\footnote{Labeled as eAdS-C in Fig.~\ref{fig2}.}) \cite{Concha:2023bly}, whose flat limit gives rise to an extended Carroll algebra. On the other hand, two successive expansions lead to an extended AdS-Static algebra\footnote{Also denoted as general Para-Bargmann in \cite{Concha:2023bly} and labeled as eAdS-S in Fig.~\ref{fig2}.}. Each of these extended kinematical algebras, whose explicit commutation relations are listed in Appendix \ref{sec:app}, admits a non-degenerate invariant tensor, which in turn allows for the construction of a well-defined gravitational CS action. 

\begin{figure}[h!]
\centering
\includegraphics[width=.43\columnwidth]{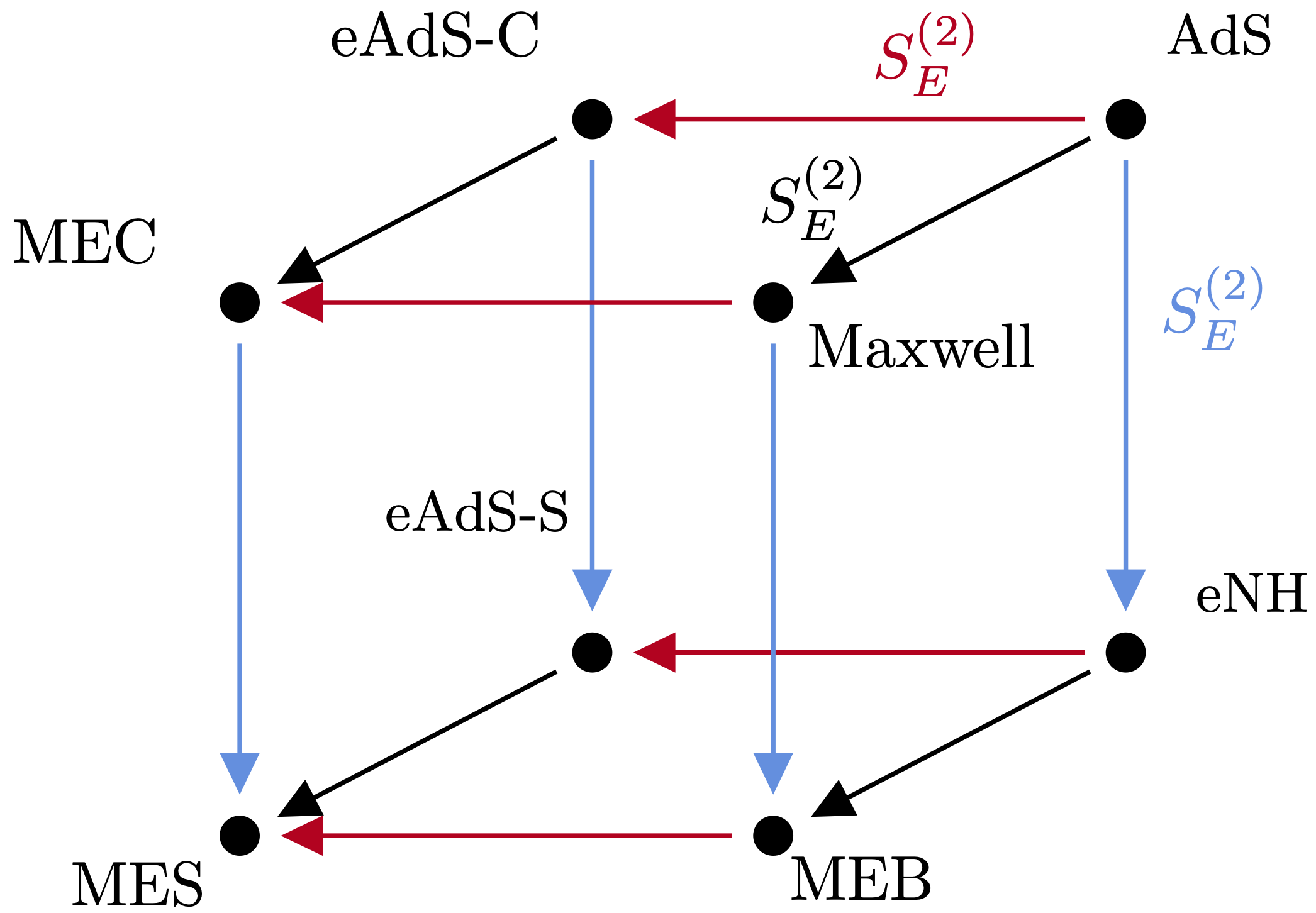}
\captionsetup{font=footnotesize}
\caption{Maxwellian generalization of the extended kinematical algebras.}
\label{fig3}
\end{figure}

The Maxwellian generalization can then be obtained by promoting the flat limit $\Lambda\rightarrow 0$ to an expansion procedure (see Fig.~\ref{fig3}). In particular, the Maxwell kinematical algebras arise from the application of a resonant $S_{E}^{(2)}$-expansion followed by a $0_S$-reduction. Remarkably, as illustrated by the cube in Fig.~\ref{fig3}, the non-Lorentzian Maxwellian algebras can also be recovered from the relativistic Maxwell algebra, inheriting the expansion relations already present in the corresponding extended kinematical algebras. 

\subsubsection*{Maxwell algebra}
Let us start with a subspace decomposition of the AdS algebra \eqref{aAdS} defined by
\begin{align}
    V_{0}&=\{J,G_{a}\}\,, & V_{1}&=\{H,P_a\}\,,
\end{align}
which satisfies a $\mathbb{Z}_2$ gradation of the form \eqref{decomp}. The Maxwell algebra is then obtained by performing a resonant $S_{E}^{(2)}$-expansion of the AdS algebra, followed by the application of the $0_S$-reduction. The Maxwell generators are related to those of $\mathfrak{so}\left(2,2\right)$ through the semigroup elements, as in Table \ref{Table1}.

\renewcommand{\arraystretch}{1.1}
\begin{table}[h!]
\centering
    \begin{tabular}{l||C{1,6cm}|C{1,6cm}|}
$\lambda_3$ & \cellcolor[gray]{0.8} & \cellcolor[gray]{0.8}  \\ \hline
$\lambda_2$ & $\mathtt{Z}$,\ \ $\mathtt{Z}_{a}$ & \cellcolor[gray]{0.8}  \\ \hline
$\lambda_1$ & \cellcolor[gray]{0.8} & $\mathtt{H}$,\ \, $\mathtt{P}_a$  \\ \hline
$\lambda_0$ & $ \mathtt{J}$,\ \ $\mathtt{G}_{a}$ & \cellcolor[gray]{0.8}  \\ \hline
 & $J$, \ $G_a$ & $H$, $P_a$ 
\end{tabular}
 \captionsetup{font=footnotesize}
\caption{Maxwell generators in terms of the AdS ones and the semigroup elements.}
\label{Table1}%
\end{table}
The commutation relations of the Maxwell algebra are obtained by combining the $\mathfrak{so}\left(2,2\right)$ commutators and the multiplication law of the semigroup $S_{E}^{(2)}$ \eqref{MLSE2},
\begin{align}
    \left[\mathtt{J},\mathtt{G}_{a}\right]&=\epsilon_{ab}\mathtt{G}_{b}\,, & \left[\mathtt{H},\mathtt{G}_{a}\right]&=\epsilon_{ab}\mathtt{P}_{b}\,, & \left[\mathtt{G}_{a},\mathtt{G}_{b}\right]&=-\epsilon_{ab}\mathtt{J}\,, \notag \\
    \left[\mathtt{J},\mathtt{P}_{a}\right]&=\epsilon_{ab}\mathtt{P}_{b}\,, & \left[\mathtt{H},\mathtt{P}_{a}\right]&=\epsilon_{ab}\mathtt{Z}_{b}\,, & \left[\mathtt{G}_{a},\mathtt{P}_{b}\right]&=-\epsilon_{ab}\mathtt{H}\,, \notag \\
    \left[\mathtt{J},\mathtt{Z}_{a}\right]&=\epsilon_{ab}\mathtt{Z}_{b}\,, & \left[\mathtt{Z},\mathtt{G}_{a}\right]&=\epsilon_{ab}\mathtt{Z}_{b}\,, & \left[\mathtt{P}_{a},\mathtt{P}_{b}\right]&=-\epsilon_{ab}\mathtt{Z}\,, \notag \\
    \left[\mathtt{G}_{a},\mathtt{Z}_{b}\right]&=-\epsilon_{ab}\mathtt{Z}\,. \label{Max}
\end{align}
It is worth noting that the Maxwell algebra \eqref{Max} is isomorphic to the extended AdS-Carroll algebra (see Table \ref{TableA2}), an isomorphism that becomes manifest upon the following identification of generators\footnote{This isomorphism motivates the denomination \textit{Para-Maxwell} algebra for the extended AdS-Carroll algebra.}:
\begin{align}
    \mathtt{G}_{a}&\leftrightarrow P_{a}\,, & \mathtt{P}_{a}&\leftrightarrow G_{a}\,, & \mathtt{Z}&\leftrightarrow C\,, & \mathtt{Z}_{a} &\leftrightarrow T_{a}\,.
\end{align}
A CS gravity action based on the Maxwell algebra can then be written in terms of a gauge connection one-form,
\begin{eqnarray}
A&=W\mathtt{J}+V\mathtt{H}+W^{a}\mathtt{G}_{a}+V^{a}\mathtt{P}_{a}+K\mathtt{Z}+K^{a}\mathtt{Z}_{a}\,,\label{Max1F}
\end{eqnarray}
where $W$ is the spin-connection for boosts, $W^{a}$ represents the spatial spin-connection, $V$ corresponds to the time-like vielbein, $V^{a}$ is the spatial vielbein, $K$ represents the time-like gravitational Maxwell field and $K^{a}$ is the spatial gravitational Maxwell field. The Maxwell algebra admits a non-degenerate invariant tensor whose non-vanishing components read
\begin{align}
    \langle \mathtt{J}\mathtt{J}\rangle&=-\alpha_0\,,& \langle\mathtt{G}_{a}\mathtt{G}_{b}\rangle&=\alpha_{0}\delta_{ab}\,, & \langle \mathtt{J}\mathtt{H}\rangle&=-\alpha_1\,,\notag\\
    \langle \mathtt{J}\mathtt{Z}\rangle&=-\alpha_2\,,& \langle\mathtt{G}_{a}\mathtt{P}_{b}\rangle&=\alpha_{1}\delta_{ab}\,, & \langle \mathtt{H}\mathtt{H}\rangle&=-\alpha_2\,,\notag\\
    \langle\mathtt{G}_{a}\mathtt{Z}_{b}\rangle&=\alpha_{2}\delta_{ab}\,, & \langle\mathtt{P}_{a}\mathtt{P}_{b}\rangle&=\alpha_{2}\delta_{ab}\,, & \label{ITmax}
\end{align}
where the non-degeneracy condition requires $\alpha_{2}\neq 0$. The CS action for the Maxwell algebra \eqref{Max} is then obtained by substituting the one-form gauge connection \eqref{Max1F} and the non-vanishing components of the invariant tensor \eqref{ITmax} into the general CS expression \eqref{csgeneral}:
\begin{eqnarray}
    I_{\text{Max}}&=&\frac{k}{4\pi}\int\alpha_{0}\left(-W dW+W_{a}R^{a}\left(W^{b}\right)\right)+\alpha_{1}\left(2V_{a}R^{a}\left(W^{b}\right)-2VR\left(W\right)\right)\notag\\
    &&+\alpha_2\left(2K_{a}R^{a}\left(W^{b}\right)+V_{a}R^{a}\left(V^{b}\right)-VR\left(V\right)-2K R\left(W\right)\right)\,. \label{MaxCS}
\end{eqnarray}
Here, $k=1/4G$ is the level of the theory related to the gravitational constant $G$ and
\begin{align}
    R\left(W\right)&=dW+\frac{1}{2}\epsilon^{ab}W_{a}W_{b}\,,\notag\\
    R^{a}\left(W^{b}\right)&=dW^{a}+\epsilon^{ab}WW_{b}\,, \notag\\
    R\left(V\right)&=dV+\epsilon^{ab}W_{a}V_{b}\,,\notag\\
    R^{a}\left(V^{b}\right)&=dV^{a}+\epsilon^{ab}W V_{b}+\epsilon^{ab}V W_{b}\,.\label{MaxC1}
\end{align}
The non-degeneracy ensures that the field equations are given by the vanishing of the curvature two-forms \eqref{MaxC1} as well as those related to the gravitational Maxwell field,
\begin{align}
    R\left(K\right)&=dK+\frac{1}{2}\epsilon^{ab}W_{a}K_{b}+\frac{1}{2}\epsilon^{ab}V_{a}V_{b}\,,\notag\\
    R^{a}\left(K^{b}\right)&=dK^{a}+\epsilon^{ab}W K_{b}+\epsilon^{ab}V V_{b}+\epsilon^{ab}K W_{b}\,.\label{MaxC2}
\end{align}
While the vanishing of the curvatures \eqref{MaxC1} describes a locally flat Riemannian geometry, the conditions imposed by the vanishing of the curvatures \eqref{MaxC2} lead to non-trivial effects when compared to General Relativity, which have been extensively studied in \cite{Salgado:2014jka,Hoseinzadeh:2014bla,Concha:2018zeb,Concha:2023nou,Aviles:2025nah}. A cosmological constant can be included in the Maxwell CS gravity theory but requires a deformation of the algebra to the so-called AdS-Lorentz algebra \cite{Soroka:2006aj,Salgado:2014qqa,Concha:2018jjj}. 

As a final remark, it is worth noticing that the CS action \eqref{MaxCS} can be recovered from the $\mathfrak{so}\left(2,2\right)$ CS action,
\begin{eqnarray}
    I_{\text{AdS}}&=&\frac{k}{4\pi}\int\mu_{0}\left(-W dW+W_{a}R^{a}\left(W^{b}\right)+V_{a}R^{a}\left(V^{b}\right)-VR\left(V\right)\right)\notag\\
    &&+\mu_{1}\left(2V_{a}R^{a}\left(W^{b}\right)+\epsilon^{ab}VV_{a}V_{b}-2VR\left(W\right)\right)\,. \label{AdSCS}
\end{eqnarray}
Indeed, the expansion procedure can be made by identifying the Maxwell gauge fields in terms of the $\mathfrak{so}\left(2,2\right)$ ones through the semigroup elements as
\begin{align}
    W&=\lambda_0 W\,, & W^{a}&=\lambda_0 W^{a}\,, & V&=\lambda_{1}V\,, \notag\\
    V^{a}&=\lambda_1 V^{a}\,, & K&=\lambda_2 W\,, & K^{a}&=\lambda_{2}W^{a}\,, 
\end{align}
and expressing the $\alpha$'s constant in terms of the $\mathfrak{so}\left(2,2\right)$ ones as
\begin{align}
    \alpha_0&=\lambda_{0}\mu_0\,, & \alpha_1&=\lambda_{1}\mu_{1}\,, & \alpha_{2}&=\lambda_{2}\mu_0\,.
\end{align}
\subsubsection*{Maxwellian extended Bargmann algebra}
Here we show that the Maxwellian version of the extended Bargmann (MEB) algebra \cite{Aviles:2018jzw} admits two distinct origins within the $S$-expansion framework. In particular, it can be obtained either from the extended Newton-Hooke algebra (see Table~\ref{TableA1}) or, alternatively, from the Maxwell algebra \eqref{Max}, as summarized in Fig.~\ref{fig3}. As in the previous case, the S-expansion requires a decomposition into subspaces (see Table~\ref{Table2}) satisfying a $\mathbb{Z}_2$-graded Lie algebra structure of the form \eqref{decomp}.

\begin{table}[h]
    \centering
    \begin{tabular}{|c||C{6cm}|C{6cm}|}
    \hline
     \rowcolor[gray]{0.9}  Subspaces  & Extended Newton-Hooke origin & Maxwell origin \\ \hline
        $V_0$ & $J,G_a,S$& $\mathtt{J},\mathtt{H},\mathtt{Z}$ \\
       \rowcolor[gray]{0.9} $V_1$ & $H,P_{a},M$ & $\mathtt{G}_{a}, \mathtt{P}_{a}, \mathtt{Z}_{a}$ \\
     \hline
         \end{tabular}
         \captionsetup{font=footnotesize}
    \caption{Subspaces decomposition of the extended Newton-Hooke and Maxwell algebra.}
    \label{Table2}
\end{table}
The commutation relations of the MEB algebra are then obtained by considering a resonant $S_{E}^{(2)}$-expansion, followed by the corresponding $0_{S}$-reduction. In particular, the MEB generators are related to those of the extended Newton-Hooke or Maxwell algebras through the semigroup elements, as shown in Table~\ref{Table3}.
\renewcommand{\arraystretch}{1.2}
\begin{table}[h!]
\centering
    \begin{tabular}{l||C{2.5cm}|C{2.5cm}||C{2.5cm}|C{2.5cm}||}
& \multicolumn{2}{|c||}{Extended Newton-Hooke origin}& \multicolumn{2}{|c||}{Maxwell origin}\\ \hline
$\lambda_3$ & \cellcolor[gray]{0.8} & \cellcolor[gray]{0.8} & \cellcolor[gray]{0.8} & \cellcolor[gray]{0.8}  \\ \hline
$\lambda_2$ & $\mathtt{Z}$,\ \ \, $\mathtt{Z}_{a}$, \ \,$\mathtt{T}$  & \cellcolor[gray]{0.8} & $\mathtt{S}$, \ \ $\mathtt{M}$, \ \ $\mathtt{T}$ & \cellcolor[gray]{0.8} \\ \hline
$\lambda_1$ & \cellcolor[gray]{0.8} & $\mathtt{H}$,\ \, $\mathtt{P}_a$,\ \, $\mathtt{M}$ & \cellcolor[gray]{0.8} & $\mathtt{G}_{a}$, \ $\mathtt{P}_{a}$, \ $\mathtt{Z}_{a}$ \\ \hline
$\lambda_0$ & $ \mathtt{J}$,\ \  \,$\mathtt{G}_{a}$, \ \ $ \mathtt{S}$ & \cellcolor[gray]{0.8} & $\mathtt{J}$, \ \ $\mathtt{H}$, \ \ $\mathtt{Z}$ & \cellcolor[gray]{0.8} \\ \hline
 & $J$, \ $G_a$, \ $S$ & $H$,\ \,$P_a$,\ \ $M$ & $\mathtt{J}$, \ \ $\mathtt{H}$, \ \ $\mathtt{Z}$ & $\mathtt{G}_{a}$, \ $\mathtt{P}_{a}$, \ $\mathtt{Z}_{a}$
\end{tabular}
 \captionsetup{font=footnotesize}
 \caption{MEB generators expressed in terms of the generators of the extended Newton-Hooke and Maxwell algebras through the $S_E^{(2)}$ semigroup elements.}
\label{Table3}%
\end{table}
Then, the commutators of the MEB algebra can be obtained by combining either the extended Newton-Hooke or the Maxwell commutators with the multiplication law of the semigroup $S_{E}^{(2)}$ \eqref{MLSE2},
\begin{align}
    \left[\mathtt{J},\mathtt{G}_{a}\right]&=\epsilon_{ab}\mathtt{G}_{b}\,, & \left[\mathtt{H},\mathtt{G}_{a}\right]&=\epsilon_{ab}\mathtt{P}_{b}\,, & \left[\mathtt{G}_{a},\mathtt{G}_{b}\right]&=-\epsilon_{ab}\mathtt{S}\,, \notag \\
    \left[\mathtt{J},\mathtt{P}_{a}\right]&=\epsilon_{ab}\mathtt{P}_{b}\,, & \left[\mathtt{H},\mathtt{P}_{a}\right]&=\epsilon_{ab}\mathtt{Z}_{b}\,, & \left[\mathtt{G}_{a},\mathtt{P}_{b}\right]&=-\epsilon_{ab}\mathtt{M}\,, \notag \\
    \left[\mathtt{J},\mathtt{Z}_{a}\right]&=\epsilon_{ab}\mathtt{Z}_{b}\,, & \left[\mathtt{Z},\mathtt{G}_{a}\right]&=\epsilon_{ab}\mathtt{Z}_{b}\,, & \left[\mathtt{P}_{a},\mathtt{P}_{b}\right]&=-\epsilon_{ab}\mathtt{T}\,, \notag \\
    \left[\mathtt{G}_{a},\mathtt{Z}_{b}\right]&=-\epsilon_{ab}\mathtt{T}\,. \label{MEB}
\end{align}
This construction shows that the MEB algebra can be understood either as a Maxwell-type extension of the extended Newton-Hooke algebra or, alternatively, as a non-relativistic expansion of the Maxwell algebra itself\footnote{The present construction corresponds to a finite subcase of the infinite-dimensional Galilean expansion of the Maxwell algebra discussed in \cite{Gomis:2019nih}.}, as summarized in Table~\ref{Table2}. The presence of the additional generators $\{\mathtt{Z}, \mathtt{Z}_a, \mathtt{T}\}$ reflects the non-trivial commutator structure induced by the semigroup. The MEB algebra, unlike the extended Bargmann one (see Table \ref{TableA1}), cannot be recovered as an extension of the Galilei algebra \eqref{Gal}. Interestingly, the MEB algebra \eqref{MEB} is isomorphic to the extended AdS-Static algebra (see Table \ref{TableA2}) upon the following identification of the generators:
\begin{align}
    \mathtt{G}_{a}&\leftrightarrow P_{a}\,, & \mathtt{P}_{a}&\leftrightarrow G_{a}\,, & \mathtt{Z}&\leftrightarrow C\,, & \mathtt{T}&\leftrightarrow B\,, & \mathtt{Z}_{a} &\leftrightarrow T_{a}\,.
\end{align}
This observation suggests relabeling the extended AdS-Static algebra as the \textit{Para-MEB} algebra. The MEB algebra admits the following non-vanishing components of the invariant tensor:
\begin{align}
    \langle \mathtt{J}\mathtt{S}\rangle&=-\beta_0\,,& \langle\mathtt{G}_{a}\mathtt{G}_{b}\rangle&=\beta_{0}\delta_{ab}\,, & \langle \mathtt{J}\mathtt{M}\rangle&=-\beta_1\,,\notag\\
    \langle \mathtt{J}\mathtt{T}\rangle&=-\beta_2\,,& \langle\mathtt{G}_{a}\mathtt{P}_{b}\rangle&=\beta_{1}\delta_{ab}\,, & \langle \mathtt{H}\mathtt{S}\rangle&=-\beta_1\,,\notag\\
    \langle \mathtt{S}\mathtt{Z}\rangle&=-\beta_2\, & \langle\mathtt{P}_{a}\mathtt{P}_{b}\rangle&=\beta_{2}\delta_{ab}\,, & \langle \mathtt{H}\mathtt{M}\rangle&=-\beta_2\,, \notag \\
    \langle\mathtt{G}_{a}\mathtt{Z}_{b}\rangle&=\beta_{2}\delta_{ab}\,, \label{ITmeb}
\end{align}
where the non-degeneracy condition of the invariant tensor requires $\beta_{2}\neq 0$. The gauge connection one-form for the MEB algebra reads
\begin{eqnarray}
    A&=&\omega \mathtt{J}+\omega^{a}\mathtt{G}_{a}+\tau \mathtt{H}+e^{a}\mathtt{P}_{a}+k\mathtt{Z}+k^{a}\mathtt{Z}_{a}+s\mathtt{S}+m\mathtt{M}+t\mathtt{T}\,.\label{1FMEB}
\end{eqnarray}
The CS gravity action for the Maxwell algebra \cite{Aviles:2018jzw} is obtained by inserting the gauge connection one-form \eqref{1FMEB} and the invariant tensor \eqref{ITmeb} into the general CS expression \eqref{csgeneral},
\begin{align}
    I_{\text{MEB}}=&\frac{k}{4\pi}\int\beta_{0}\left(-s R\left(\omega\right)+\omega_{a}R^{a}\left(\omega^{b}\right)\right)+\beta_{1}\left(2e_{a}R^{a}\left(\omega^{b}\right)-2mR\left(\omega\right)-2\tau R\left(s\right)\right)\notag\\
&+\beta_2\left(k_{a}R^{a}\left(\omega^{b}\right)+\omega^{a}R_{a}\left(k^{b}\right)+e_{a}R^{a}\left(e^{b}\right)-2s R\left(k\right)-2mR\left(\tau\right)-2t R\left(\omega\right)\right)\,, \label{MEBCS}
\end{align}
where
\begin{align}
    R\left(\omega\right)&=d\omega\,, &
    R^{a}\left(\omega^{b}\right)&=d\omega^{a}+\epsilon^{ab}\omega \omega_{b}\,, \notag\\
    R\left(\tau\right)&=d\tau\,, &
    R^{a}\left(e^{b}\right)&=de^{a}+\epsilon^{ab}\omega e_{b}+\epsilon^{ab}\tau \omega_{b}\,,\notag\\
    R\left(k\right)&=dk\,, &
    R^{a}\left(k^{b}\right)&=dk^{a}+\epsilon^{ab}\omega k_{b}+\epsilon^{ab} \tau e_{b}+\epsilon^{ab}k \omega_{b}\,,\notag\\
    R\left(s\right)&=ds+\frac{1}{2}\epsilon^{ab} \omega_{a}\omega_{b}\,. \label{CurvMEB1}
\end{align}
It is worth emphasizing that the first two sectors, proportional to $\beta_0$ and $\beta_{1}$, reproduce the most general CS action for the extended Bargmann algebra \cite{Penafiel:2019czp,Concha:2020eam}. The Maxwellian contribution appears explicitly in the $\beta_2$ sector. As its relativistic version, a cosmological constant can be included deforming the symmetry algebra to an enlarged extended Bargmann algebra \cite{Concha:2019lhn}. The field equations, for $\beta_{2}\neq 0$, are given by the vanishing of the MEB curvature two-forms represented by \eqref{CurvMEB1} and
\begin{align}
    R\left(m\right)&=dm+\epsilon^{ab}\omega_{a}e_{b}\,, & R\left(t\right)&=dt+\epsilon^{ab}\omega_{a}k_{b}+\frac{1}{2}\epsilon^{ab}e_{a}e_{b}\,.
\end{align}
The CS action \eqref{MEBCS} for the MEB algebra can alternatively be derived from the most general extended Newton-Hooke CS action \cite{Penafiel:2019czp,Concha:2020eam},
\begin{eqnarray}
    I_{\text{eNH}}&=&\frac{k}{4\pi}\int\nu_{0}\left(-s R\left(\omega\right)+\omega_{a}R^{a}\left(\omega^{b}\right)+e_{a}R^{a}\left(e^{b}\right)-2mR\left(\tau\right)\right)\notag\\
    &&+\nu_{1}\left(2e_{a}R^{a}\left(\omega^{b}\right)-2mR\left(\omega\right)-2\tau R\left(s\right)+\epsilon^{ab}\tau e_{a}e_{b}\right)\,, \label{eNHCS}
\end{eqnarray}
by expressing the MEB gauge fields in terms of the extended Newton-Hooke ones and the semigroup elements as
\begin{align}
    \omega&=\lambda_{0}\omega\,, & \omega^{a}&=\lambda_{0}\omega^{a}\,, & s&=\lambda_0 s\,,\notag \\
    \tau&=\lambda_{1} \tau\,, & e^{a}&=\lambda_{1}e^{a}\,, & m&=\lambda_{1} m\,,\notag\\
    k&=\lambda_{2}\omega\,, & k^{a}&=\lambda_2\omega^{a}\,, & t&=\lambda_{2} s\,,
\end{align}
and identifying the $\beta$'s constant as
\begin{align}
    \beta_0&=\lambda_{0}\nu_0\,, & \beta_1&=\lambda_{1}\nu_{1}\,, & \beta_{2}&=\lambda_{2}\nu_0\,.
\end{align}
Similarly, the MEB CS action \eqref{MEBCS} can be obtained from the Maxwell CS gravity action \eqref{MaxCS} by expressing the MEB gauge fields in terms of the Maxwell ones through the $S_E^{(2)}$ semigroup elements, according to the identifications of their respective generators summarized in Table~\ref{Table3}.
\subsubsection*{Maxwellian extended Carroll algebra}
At the ultra-relativistic level, we show that the non-degenerate Maxwellian extension of the Carroll algebra, denoted as the Maxwellian extended Carroll (MEC) algebra \cite{Concha:2021jnn}, can be obtained either from the extended AdS-Carroll algebra (see Table~\ref{TableA2}) or from the Maxwell algebra \eqref{Max}, as illustrated in Fig.~\ref{fig3}. The MEC algebra is obtained by performing a resonant $S_{E}^{(2)}$-expansion of the starting algebra followed by a $0_S$-reduction. To this end, we first consider the subspace decomposition shown in Table~\ref{Table4}.

\begin{table}[h]
    \centering
    \begin{tabular}{|c||C{6cm}|C{6cm}|}
    \hline
     \rowcolor[gray]{0.9}  Subspaces  & Extended AdS-Carroll origin & Maxwell origin \\ \hline
        $V_0$ & $J,G_a,C$& $\mathtt{J},\mathtt{P}_{a},\mathtt{Z}$ \\
       \rowcolor[gray]{0.9} $V_1$ & $H,P_{a},T_{a}$ & $\mathtt{G}_{a}, \mathtt{H}, \mathtt{Z}_{a}$ \\
     \hline
         \end{tabular}
         \captionsetup{font=footnotesize}
    \caption{Subspaces decomposition of the extended AdS-Carroll and Maxwell algebra.}
    \label{Table4}
\end{table}
The MEC generators can be recovered from the extended AdS-Carroll or the Maxwell generators through the semigroup elements, as displayed in Table \ref{Table5}.

\renewcommand{\arraystretch}{1.2}
\begin{table}[h!]
\centering
    \begin{tabular}{l||C{2.5cm}|C{2.5cm}||C{2.5cm}|C{2.5cm}||}
& \multicolumn{2}{|c||}{Extended AdS-Carroll origin}& \multicolumn{2}{|c||}{Maxwell origin}\\ \hline
$\lambda_3$ & \cellcolor[gray]{0.8} & \cellcolor[gray]{0.8} & \cellcolor[gray]{0.8} & \cellcolor[gray]{0.8} \\ \hline
$\lambda_2$ & $\mathtt{Z}$,\ \ \,$\mathtt{Z}_{a}$, \ \,$\mathtt{L}$  & \cellcolor[gray]{0.8} & $\mathtt{C}$, \ \ $\mathtt{T}_{a}$, \ \ $\mathtt{L}$ & \cellcolor[gray]{0.8} \\ \hline
$\lambda_1$ & \cellcolor[gray]{0.8} & $\mathtt{H}$,\ \, $\mathtt{P}_a$,\ \, $\mathtt{T}_{a}$ & \cellcolor[gray]{0.8} & $\mathtt{G}_{a}$, \ \ $\mathtt{H}$, \ \ $\mathtt{Z}_{a}$ \\ \hline
$\lambda_0$ & $\mathtt{J}$,\ \  \,$\mathtt{G}_{a}$, \ \ $ \mathtt{C}$ & \cellcolor[gray]{0.8} & $\mathtt{J}$, \ \ $\mathtt{P}_a$, \ \ $\mathtt{Z}$ & \cellcolor[gray]{0.8} \\ \hline
 & $J$, \ $G_a$, \ $C$ & $H$,\ \,$P_a$,\ \ $T_{a}$ & $\mathtt{J}$, \ \ $\mathtt{P}_a$, \ \ $\mathtt{Z}$  & $\mathtt{G}_{a}$, \ \ $\mathtt{H}$, \ \ $\mathtt{Z}_{a}$
\end{tabular}
 \captionsetup{font=footnotesize}
\caption{MEC generators expressed in terms of the extended AdS-Carroll and Maxwell generators through the $S_E^{(2)}$ semigroup elements.}
\label{Table5}%
\end{table}
The MEC algebra satisfies the following commutation relations:
\begin{align}
    \left[\mathtt{J},\mathtt{G}_{a}\right]&=\epsilon_{ab}\mathtt{G}_{b}\,, & \left[\mathtt{Z},\mathtt{G}_{a}\right]&=\epsilon_{ab}\mathtt{Z}_{b}\,, & \left[\mathtt{P}_{a},\mathtt{P}_{b}\right]&=-\epsilon_{ab}\mathtt{Z}\,, \notag \\
    \left[\mathtt{J},\mathtt{P}_{a}\right]&=\epsilon_{ab}\mathtt{P}_{b}\,, & \left[\mathtt{H},\mathtt{P}_{a}\right]&=\epsilon_{ab}\mathtt{Z}_{b}\,, & \left[\mathtt{G}_{a},\mathtt{P}_{b}\right]&=-\epsilon_{ab}\mathtt{H}\,, \notag \\
    \left[\mathtt{J},\mathtt{Z}_{a}\right]&=\epsilon_{ab}\mathtt{Z}_{b}\,, & \left[\mathtt{H},\mathtt{G}_{a}\right]&=\epsilon_{ab}\mathtt{T}_{b}\,, & \left[\mathtt{G}_{a},\mathtt{G}_{b}\right]&=-\epsilon_{ab}\mathtt{C}\,, \notag \\
    \left[\mathtt{J},\mathtt{T}_{a}\right]&=\epsilon_{ab}\mathtt{T}_{b}\,, & \left[\mathtt{C},\mathtt{P}_{a}\right]&=\epsilon_{ab}\mathtt{T}_{b}\,,& \left[\mathtt{G}_{a},\mathtt{Z}_{b}\right]&=-\epsilon_{ab}\mathtt{L}\,,\notag \\
    \left[\mathtt{P}_{a},\mathtt{T}_{b}\right]&=-\epsilon_{ab}\mathtt{L}\,.\label{MEC}
\end{align}
These commutation relations are obtained by combining either the extended AdS-Carroll or the Maxwell commutators with the multiplication law of the semigroup $S_{E}^{(2)}$ given in \eqref{MLSE2}. Let us note that the MEC algebra, which contains 13 generators, is not isomorphic to the MEB algebra \eqref{MEB}, which is spanned by 12 generators. In particular, the MEC algebra is characterized by the presence of a central charge $\mathtt{L}$, whose presence ensures the non-degenercy of the invariant tensor, which reads
\begin{align}
    \langle \mathtt{J}\mathtt{J}\rangle&=-\gamma_0\,,& \langle\mathtt{G}_{a}\mathtt{G}_{b}\rangle&=\gamma_{0}\delta_{ab}\,, & \langle \mathtt{J}\mathtt{C}\rangle&=-\gamma_0\,,\notag\\
    \langle \mathtt{J}\mathtt{H}\rangle&=-\gamma_1\,,& \langle\mathtt{G}_{a}\mathtt{P}_{b}\rangle&=\gamma_{1}\delta_{ab}\,, & \langle \mathtt{J}\mathtt{Z}\rangle&=-\gamma_2\,, \notag\\
    \langle \mathtt{J}\mathtt{L}\rangle&=-\gamma_2\,,& \langle\mathtt{P}_{a}\mathtt{P}_{b}\rangle&=\gamma_{2}\delta_{ab}\,, & \langle \mathtt{H}\mathtt{H}\rangle&=-\gamma_2\,,\notag\\
    \langle \mathtt{C}\mathtt{Z}\rangle&=-\gamma_2\,, & \langle\mathtt{G}_{a}\mathtt{Z}_{b}\rangle&=\gamma_{2}\delta_{ab}\,,&  \langle\mathtt{P}_{a}\mathtt{T}_{b}\rangle&=\gamma_{2}\delta_{ab}\,.
     \label{ITmec}
\end{align}
The MEC CS gravity action is written in terms of the gauge connection one-form,
\begin{eqnarray}
    A&=&\omega \mathtt{J}+\omega^{a}\mathtt{G}_{a}+\tau \mathtt{H}+e^{a}\mathtt{P}_{a}+k\mathtt{Z}+k^{a}\mathtt{Z}_{a}+\varsigma\mathtt{C}+t^{a}\mathtt{T}_{a}+l\mathtt{L}\,.\label{1FMEC}
\end{eqnarray}
The MEC CS action, first introduced in \cite{Concha:2021jnn}, reads
\begin{align}
    I_{\text{MEC}}=\frac{k}{4\pi}&\int\gamma_{0}\left(-\omega R\left(\omega\right)+\omega_{a}R^{a}\left(\omega^{b}\right)-2\varsigma R\left(\omega\right)\right)+\gamma_{1}\left(2e_{a}R^{a}\left(\omega^{b}\right)-2\tau R\left(\omega\right)\right)\notag\\
&+\gamma_2\left(k_{a}R^{a}\left(\omega^{b}\right)+\omega^{a}R_{a}\left(k^{b}\right)+e_{a}R^{a}\left(e^{b}\right)-2\varsigma R\left(k\right)-2l R\left(\omega\right)-2kR\left(\omega\right)\right.\notag\\
&\left.+2t_{a}R\left(e^{b}\right)-\tau R\left(\tau\right)\right)\,, \label{MECCS}
\end{align}
where
\begin{align}
    R\left(\omega\right)&=d\omega\,, &
    R^{a}\left(\omega^{b}\right)&=d\omega^{a}+\epsilon^{ab}\omega \omega_{b}\,, \notag\\
    R\left(\tau\right)&=d\tau+\epsilon^{ab}\omega_{a}e_{b}\,, &
    R^{a}\left(e^{b}\right)&=de^{a}+\epsilon^{ab}\omega e_{b}\,,\notag\\
    R\left(k\right)&=dk+\frac{1}{2}\epsilon^{ab}e_{a}e_{b}\,, &
    R^{a}\left(k^{b}\right)&=dk^{a}+\epsilon^{ab}\omega k_{b}+\epsilon^{ab} \tau e_{b}+\epsilon^{ab}k \omega_{b}\,.\label{CurvMEC1}
\end{align}
The non-degeneracy of the invariant tensor \eqref{ITmec} ensures that the field equations are given by the vanishing of the MEC curvature two-forms represented by \eqref{CurvMEC1} and 
\begin{align}
   R\left(\varsigma\right)&=d\varsigma+\frac{1}{2}\epsilon^{ab} \omega_{a}\omega_{b}\,, \notag\\
R\left(l\right)&=dl+\epsilon^{ab}\omega_{a}k_{b}+\epsilon^{ab}e_{a}t_{b}\,,\notag\\
    R^{a}\left(t^{b}\right)&=dt^{a}+\epsilon^{ab}\omega t_{b}+\epsilon^{ab}\varsigma e_{b}+\epsilon^{ab}\tau \omega_{b}\,.
\end{align}
As in the original Carroll algebra, the equations of motion are characterized by a non-vanishing temporal torsion,
\begin{equation}
    d\tau=-\epsilon^{ab}\omega_{a}e_{b}\,.
\end{equation}
Nonetheless, the Maxwellian extension introduces additional gauge fields $\{\varsigma,t^{a},l\}$, associated with the generators $\{\mathtt{C},\mathtt{T}_{a},\mathtt{L}\}$, which are required in order to ensure a non-degenerate invariant tensor and thus a well-defined set of field equations. While the role of Maxwell gauge fields at the relativistic level has been explored in different contexts \cite{Hoseinzadeh:2014bla,Concha:2018zeb,Concha:2024rac,Aviles:2025nah}, the physical interpretation of the additional content in the Carrollian regime remains largely open.

It is worth pointing out that the MEC CS action \eqref{MECCS} can be directly obtained from the extended AdS-Carroll CS action
\begin{eqnarray}
    I_{\text{eAdS-C}}&=&\frac{k}{4\pi}\int\sigma_{0}\left(e_{a}R^{a}\left(e^{b}\right)-\omega R\left(\omega\right)\right)+\sigma_{1}\left(2e_{a}R^{a}\left(\omega^{b}\right)-2\tau R\left(\omega\right)+\epsilon^{ab}\tau e_{a}e_{b}\right) \notag\\
    &&+\sigma_{2}\left(\omega_{a}R^{a}\left(\omega^{b}\right)-2\tau R\left(\tau\right)-2\varsigma R\left(\omega\right)+e_{a}R^{a}\left(t^{b}\right)+t_{a}R^{a}\left(e^{b}\right)\right)\,,\label{eADSCCS}
\end{eqnarray}
upon implementing the field identifications induced by the $S_{E}^{(2)}$ semigroup elements,
\begin{align}
    \omega&=\lambda_{0} \omega\,, & \omega^{a}&=\lambda_{0}\omega^{a}\,, & \varsigma&=\lambda_0 \varsigma\,,\notag \\
    \tau&=\lambda_{1} \tau\,, & e^{a}&=\lambda_{0}e^{a}\,, & t^{a}&=\lambda_{1} t^{a}\,,\notag\\
    k&=\lambda_{2} \omega\,, & k^{a}&=\lambda_2 \omega^{a}\,, & l&=\lambda_{2} \varsigma\,,
\end{align}
together with the identification of the coupling constants,
\begin{align}
    \gamma_0&=\lambda_{0}\sigma_0\,, & \gamma_1&=\lambda_{1}\sigma_{1}\,, & \gamma_{2}&=\lambda_{2}\sigma_{2}\,.
\end{align}
In a similar manner, the MEC action \eqref{MECCS} can also be derived from the Maxwell CS gravity action \eqref{MaxCS} by expressing the MEC gauge fields in terms of the Maxwell ones through the $S_{E}^{(2)}$ semigroup elements, according to the generator identifications displayed in Table~\ref{Table5}.
\subsubsection*{Maxwellian extended static algebra}
For completeness, we close this section by introducing a Maxwellian version of the extended static algebra, which we denote as the MES algebra. Although this symmetry algebra has not been previously discussed in the literature, we show that the $S$-expansion framework provides three independent constructions leading to its commutation relations. In particular, the MES algebra can be obtained from the extended AdS-static, the MEB and the MEC algebras (see Table~\ref{Table6}).

\begin{table}[h]
    \centering
    \begin{tabular}{|c||C{4cm}|C{4cm}|C{4cm}|}
    \hline
     \rowcolor[gray]{0.9}  Subspaces  & Extended AdS-static & MEB & MEC \\
     \rowcolor[gray]{0.9} & origin & origin & origin \\ \hline
        $V_0$ & $J,G_a,C,S,B$& $\mathtt{J},\mathtt{P}_{a},\mathtt{Z},\mathtt{S},\mathtt{T}$& $\mathtt{J}, \mathtt{C},\mathtt{Z},\mathtt{L},\mathtt{H}$  \\
       \rowcolor[gray]{0.9} $V_1$ & $H,P_{a},T_{a},M$ & $\mathtt{G}_{a}, \mathtt{H}, \mathtt{Z}_{a}, \mathtt{M}$& $\mathtt{G}_{a},\mathtt{P}_{a},\mathtt{T}_{a},\mathtt{Z}_{a}$ \\
     \hline
         \end{tabular}
         \captionsetup{font=footnotesize}
    \caption{Subspaces decomposition of the extended AdS-static, MEB and MEC algebra.}
    \label{Table6}
\end{table}
The MES algebra arises from a resonant $S_{E}^{(2)}$-expansion of any of the starting algebras, followed by a $0_S$-reduction. The resulting MES generators are related to the generators of the parent algebras through the semigroup elements, as summarized in Table~\ref{Table7}.

\renewcommand{\arraystretch}{1.2}
\begin{table}[h!]
\centering
    \begin{tabular}{l||C{2.1cm}|C{2.1cm}||C{2.0cm}|C{1.9cm}||C{1.9cm}|C{1.9cm}||}
& \multicolumn{2}{|c||}{Extended AdS-static origin}& \multicolumn{2}{|c||}{MEB origin}&\multicolumn{2}{|c||}{MEC origin}\\ \hline
$\lambda_3$ & \cellcolor[gray]{0.8} & \cellcolor[gray]{0.8} & \cellcolor[gray]{0.8} & \cellcolor[gray]{0.8}& \cellcolor[gray]{0.8} & \cellcolor[gray]{0.8} \\ \hline
$\lambda_2$ & $\mathtt{Z}$, $\mathtt{Z}_{a}$, $\mathtt{L}$, $\mathtt{T}$, $\mathtt{Y}$  & \cellcolor[gray]{0.8} & $\mathtt{C}$,\,$\mathtt{T}_{a}$,\,$\mathtt{L}$,\,$\mathtt{B}$,\,$\mathtt{Y}$ & \cellcolor[gray]{0.8} & $\mathtt{S},\mathtt{B},\mathtt{T},\mathtt{Y},\mathtt{M}$ & \cellcolor[gray]{0.8} \\ \hline
$\lambda_1$ & \cellcolor[gray]{0.8} & $\mathtt{H}$,  $\mathtt{P}_a$, $\mathtt{T}_{a}$, $\mathtt{M}$ & \cellcolor[gray]{0.8} & $\mathtt{G}_{a}$,\,$\mathtt{H}$,\,$\mathtt{Z}_{a}$,\,$\mathtt{M}$ & \cellcolor[gray]{0.8} & $\mathtt{G}_{a},\mathtt{P}_{a},\mathtt{T}_{a},\mathtt{Z}_{a}$ \\ \hline
$\lambda_0$ & $\mathtt{J}$, $\mathtt{G}_{a}$, $\mathtt{C}$, $\mathtt{S}$,  $\mathtt{B}$ &  \cellcolor[gray]{0.8} & $\mathtt{J}$,\,$\mathtt{P}_a$,\,$\mathtt{Z}$,\,$\mathtt{S}$,\,$\mathtt{T}$ & \cellcolor[gray]{0.8} & $\mathtt{J},\mathtt{C},\mathtt{Z},\mathtt{L},\mathtt{H}$ & \cellcolor[gray]{0.8} \\ \hline
 & $J,G_a,C,S,B$ & $H,P_a,T_{a},M$& $\mathtt{J},\mathtt{P}_{a},\mathtt{Z}, \mathtt{S}, \mathtt{T}$ & $\mathtt{G}_{a},\mathtt{H},\mathtt{Z}_{a},\mathtt{M}$ & $\mathtt{J},\mathtt{C},\mathtt{Z},\mathtt{L},\mathtt{H}$ & $\mathtt{G}_{a},\mathtt{P}_{a},\mathtt{T}_{a},\mathtt{Z}_{a}$
\end{tabular}
 \captionsetup{font=footnotesize}
\caption{MES generators expressed in terms of the extended AdS-static, MEB and MEC generators through the $S_E^{(2)}$ semigroup elements.}
\label{Table7}%
\end{table}
The commutation relations of the MES algebra follow from combining the commutators of the chosen starting algebra with the multiplication law of the $S_{E}^{(2)}$ semigroup \eqref{MLSE2}:
\begin{align}
    \left[\mathtt{J},\mathtt{G}_{a}\right]&=\epsilon_{ab}\mathtt{G}_{b}\,, & \left[\mathtt{Z},\mathtt{G}_{a}\right]&=\epsilon_{ab}\mathtt{Z}_{b}\,, & \left[\mathtt{P}_{a},\mathtt{P}_{b}\right]&=-\epsilon_{ab}\mathtt{T}\,, \notag \\
    \left[\mathtt{J},\mathtt{P}_{a}\right]&=\epsilon_{ab}\mathtt{P}_{b}\,, & \left[\mathtt{H},\mathtt{P}_{a}\right]&=\epsilon_{ab}\mathtt{Z}_{b}\,, & \left[\mathtt{G}_{a},\mathtt{P}_{b}\right]&=-\epsilon_{ab}\mathtt{M}\,, \notag \\
    \left[\mathtt{J},\mathtt{Z}_{a}\right]&=\epsilon_{ab}\mathtt{Z}_{b}\,, & \left[\mathtt{H},\mathtt{G}_{a}\right]&=\epsilon_{ab}\mathtt{T}_{b}\,, & \left[\mathtt{G}_{a},\mathtt{G}_{b}\right]&=-\epsilon_{ab}\mathtt{B}\,, \notag \\
    \left[\mathtt{J},\mathtt{T}_{a}\right]&=\epsilon_{ab}\mathtt{T}_{b}\,, & \left[\mathtt{C},\mathtt{P}_{a}\right]&=\epsilon_{ab}\mathtt{T}_{b}\,,& \left[\mathtt{G}_{a},\mathtt{Z}_{b}\right]&=-\epsilon_{ab}\mathtt{Y}\,,\notag \\
    \left[\mathtt{P}_{a},\mathtt{T}_{b}\right]&=-\epsilon_{ab}\mathtt{Y}\,.\label{MES}
\end{align}
Let us note that the MES algebra contains two additional $\mathfrak{u}(1)$ central generators, $\mathtt{S}$ and $\mathtt{L}$, satisfying
\begin{equation}
    [\mathtt{X},\mathtt{S}]=[\mathtt{X},\mathtt{L}]=0\,,\qquad \forall\,\mathtt{X}\in\mathfrak{mes}\,.
\end{equation}
Although $\mathtt{S}$ and $\mathtt{L}$ do not enter in the non-trivial commutators in \eqref{MES}, their presence is essential to ensure the non-degeneracy of the invariant bilinear form. In particular, the MES algebra admits the following non-vanishing components of the invariant tensor:
\begin{align}
    \langle \mathtt{J}\mathtt{S}\rangle&=-\zeta_0\,,& \langle\mathtt{G}_{a}\mathtt{G}_{b}\rangle&=\zeta_{0}\delta_{ab}\,, & \langle \mathtt{J}\mathtt{B}\rangle&=-\zeta_0\,,\notag\\
    \langle \mathtt{S}\mathtt{C}\rangle&=-\zeta_0\,,& \langle\mathtt{G}_{a}\mathtt{P}_{b}\rangle&=\zeta_{1}\delta_{ab}\,, & \langle \mathtt{J}\mathtt{M}\rangle&=-\zeta_1\,, \notag\\
    \langle \mathtt{S}\mathtt{H}\rangle&=-\zeta_1\,,& \langle\mathtt{P}_{a}\mathtt{P}_{b}\rangle&=\zeta_{2}\delta_{ab}\,, & \langle \mathtt{J}\mathtt{T}\rangle&=-\zeta_2\,,\notag\\
    \langle \mathtt{S}\mathtt{Z}\rangle&=-\zeta_2\,, & \langle\mathtt{G}_{a}\mathtt{Z}_{b}\rangle&=\zeta_{2}\delta_{ab}\,,&   \langle \mathtt{J}\mathtt{Y}\rangle&=-\zeta_2\,,\notag\\
     \langle \mathtt{S}\mathtt{L}\rangle&=-\zeta_2\,, &
\langle\mathtt{P}_{a}\mathtt{T}_{b}\rangle&=\zeta_{2}\delta_{ab}\,, &  \langle \mathtt{H}\mathtt{M}\rangle&=-\zeta_2\,,\notag\\
 \langle \mathtt{C}\mathtt{T}\rangle&=-\zeta_2\,, &  \langle \mathtt{B}\mathtt{Z}\rangle&=-\zeta_2\,.
     \label{ITmes}
\end{align}
where non-degeneracy requires $\zeta_2\neq 0$. Although a Maxwellian static algebra without the generators 
$\{\mathtt{C},\mathtt{L},\mathtt{B},\mathtt{Y},\mathtt{T}_{a}\}$ 
still satisfies the Jacobi identity, it does not admit a non-degenerate invariant bilinear form. Therefore, the minimal content required to obtain a consistent Maxwellian extension of the static algebra, and consequently a well-defined CS action, is precisely the complete set of generators introduced here. The corresponding CS action can be constructed straightforwardly from \eqref{ITmes} and the gauge connection associated with the MES algebra. Given its lengthy structure, we refrain from presenting it explicitly. Moreover, it may be derived directly from the extended AdS-static, MEB or MEC CS gravity actions by expressing the MES gauge field in terms of the corresponding parent algebra through the $S_{E}^{(2)}$ semigroup elements, according to the generator identifications displayed in Table~\ref{Table7}.

\section{Generalized extended kinematical algebras}\label{sec4}
\begin{figure}[h!]
\centering
\includegraphics[width=.44\columnwidth]{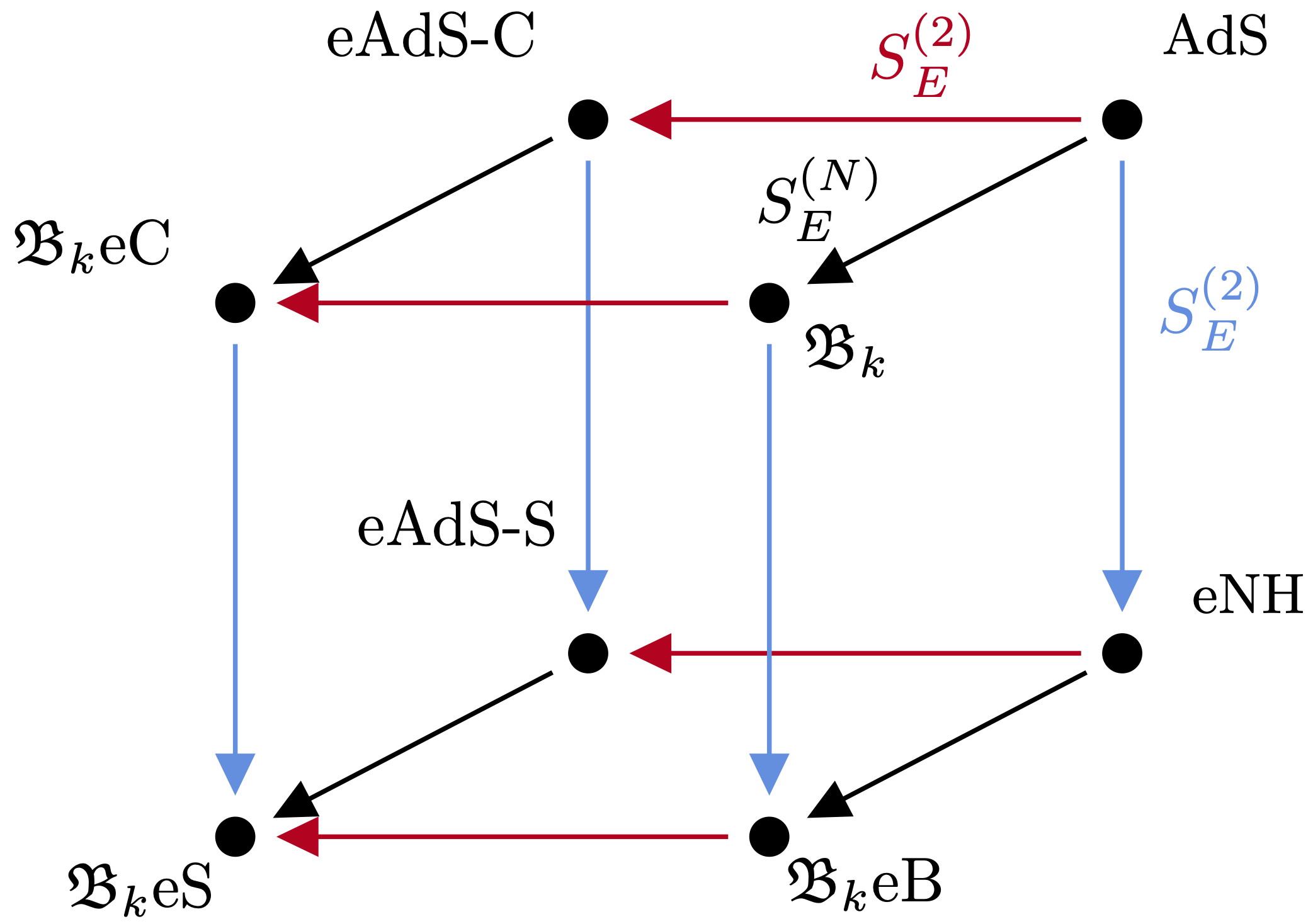}
\captionsetup{font=footnotesize}
\caption{$\mathfrak{B}_k$ generalization of the extended kinematical algebras.}
\label{fig4}
\end{figure}

In this section, we show that the original cube introduced by Bacry and Lévy-Leblond \cite{Bacry:1968zf}, together with its Maxwellian extension discussed in the previous section, admits a natural generalization through the use of an arbitrary semigroup $S_{E}^{(N)}$. This construction gives rise to an infinite hierarchy of generalized kinematical cubes associated with the non-Lorentzian sector of the so-called $\mathfrak{B}_{k}$ algebras \cite{Edelstein:2006se,Izaurieta:2009hz,Concha:2013uhq}. The $\mathfrak{B}_{k}$ generalization of the extended kinematical Lie algebras is schematically illustrated in Fig.~\ref{fig4}, where the vanishing cosmological constant limit of the the cube of Bacry and Lévy-Leblond \cite{Bacry:1968zf} is extended by means of an $S_{E}^{(N)}$-expansion with $N = k-2$. In this construction, the $S_{E}^{(2)}$-expansion is preserved along both the non-relativistic and ultra-relativistic directions in order to obtain a non-degenerate non-Lorentzian counterpart of the $\mathfrak{B}_{k}$ algebra. Higher-order semigroups along the non-relativistic or ultra-relativistic directions of the cube shown in Fig.~\ref{fig4} would instead generate post-Newtonian or post-Carrollian extensions of the $\mathfrak{B}_{k}$ algebra. Such generalizations, however, lie beyond the scope of the present work. 

Our starting point consists of the extended kinematical algebras summarized in Fig.~\ref{fig2}. We then consider, for each of these algebras, a subspace decomposition satisfying a $\mathbb{Z}_{2}$-graded Lie algebra structure, as displayed in Table~\ref{Table8}.

\begin{table}[h]
    \centering
    \begin{tabular}{|c||C{2.5cm}|C{2.5cm}|C{2.5cm}|C{2.5cm}|}
    \hline
     \rowcolor[gray]{0.9}  Subspaces  & AdS origin & eNH origin & eAdS-C origin & eAdS-S origin \\ \hline
        $V_0$ & $J,G_a$& $J,G_a,S$ & $J,G_a,C$  & $J,G_a,C,S,B$  \\
       \rowcolor[gray]{0.9} $V_1$ & $H,P_{a}$ & $H,P_{a},M$ & $H,P_{a},T_{a}$ & $H,P_{a},T_{a},M$ \\
     \hline
         \end{tabular}
         \captionsetup{font=footnotesize}
    \caption{Subspace decompositions of the extended kinematical algebras.}
    \label{Table8}
\end{table}
Let us now consider the semigroup $S_{E}^{(N)}=\{\lambda_0,\lambda_1,\cdots,\lambda_{N+1}\}$, which obeys the multiplication law
\begin{equation}
\lambda _{\alpha }\lambda _{\beta }=\left\{ 
\begin{array}{lcl}
\lambda _{\alpha +\beta }\,\ \ \, & \mathrm{if}\,\,\,\,\alpha +\beta \leq
N+1\,, &  \\ 
\lambda _{N+1}\,\,\, & \mathrm{if}\,\,\,\,\alpha +\beta >N+1\,. & 
\end{array}%
\right.   \label{MLSEN}
\end{equation}
A subset decomposition of the semigroup $S_{E}^{(N)}$, which is said to be resonant with the subspace splitting given in Table~\ref{Table8}, is defined by
\begin{align}
    S_0&=\{\lambda_{2i}\}\cup\{\lambda_{N+1}\}\,, & S_1&=\{\lambda_{2i+1}\}\cup\{\lambda_{N+1}\}\,, \label{subdecN}
\end{align}
with $i=0,1,2,\ldots,[N/2]$, where $\left[\cdot\right]$ denotes the integer part. The resonant condition then allows for the construction of a resonant $S_{E}^{(N)}$-expanded subalgebra given by \eqref{rexp}. The $\mathfrak{B}_k$ algebra and its corresponding non-Lorentzian counterparts can be obtained from the extended kinematical algebras (see Fig.~\ref{fig2}) by applying a resonant $S_{E}^{(N)}$-expansion followed by the corresponding $0_S$-reduction. The expanded generators are related to the generators of the extended kinematical algebras through the semigroup elements, as summarized in Table~\ref{Table9}.

\begin{table}[h]
    \centering
    \begin{tabular}{|c||C{2.5cm}|C{2.5cm}|C{2.5cm}|C{2.5cm}|}
    \hline
     \rowcolor[gray]{0.9}  Expanded Generators  & AdS origin & eNH origin & eAdS-C origin & eAdS-S origin \\ \hline
        $\mathtt{J}^{(m)}$ & $\lambda_{2m}J$ & $\lambda_{2m}J$ & $\lambda_{2m}J$  & $\lambda_{2m}J$  \\
       \rowcolor[gray]{0.9} $\mathtt{G}_{a}^{(m)}$ & $\lambda_{2m}G_{a}$ & $\lambda_{2m}G_{a}$ & $\lambda_{2m}G_{a}$ & $\lambda_{2m}G_{a}$ \\
       $\mathtt{S}^{(m)}$ & - & $\lambda_{2m}S$ & -  & $\lambda_{2m}S$  \\
       \rowcolor[gray]{0.9} $\mathtt{C}^{(m)}$ & - & - & $\lambda_{2m}C$  & $\lambda_{2m}C$  \\
       $\mathtt{B}^{(m)}$ & - & - & -  & $\lambda_{2m}B$  \\
       \rowcolor[gray]{0.9} $\mathtt{H}^{(m)}$ & $\lambda_{2m+1}H$ & $\lambda_{2m+1}H$ & $\lambda_{2m+1}H$  & $\lambda_{2m+1}H$  \\
       $\mathtt{P}_{a}^{(m)}$ & $\lambda_{2m+1}P_{a}$ & $\lambda_{2m+1}P_{a}$ & $\lambda_{2m+1}P_{a}$ & $\lambda_{2m+1}P_{a}$ \\
       \rowcolor[gray]{0.9} $\mathtt{M}^{(m)}$ & - & $\lambda_{2m+1}M$ & -  & $\lambda_{2m+1}M$  \\
       $\mathtt{T}_{a}^{(m)}$ & - & - & $\lambda_{2m+1}T_{a}$ & $\lambda_{2m+1}T_{a}$ \\
     \hline
         \end{tabular}
         \captionsetup{font=footnotesize}
    \caption{Expanded generators in terms of the extended kinematical ones.}
    \label{Table9}
\end{table}
By combining the original commutation relations of the extended kinematical algebras with the multiplication law of the semigroup $S_{E}^{(N)}$, one finds that the $\mathfrak{B}_k$ generalizations satisfy the commutation relations listed in Table~\ref{Table10}. This generalization shows that both the extended kinematical algebras \cite{Concha:2023bly,Concha:2024dap} and their Maxwellian extensions can be understood as particular subcases of the $\mathfrak{B}_{k}$ extended kinematical algebra. For $N=1$, the resonant $0_{S}$-reduced $S_{E}^{(N)}$-expansion reduces to a contraction corresponding to the vanishing cosmological constant limit. In this case, the expanded algebras coincide with the $\mathfrak{B}_{3}$ algebra and its non-Lorentzian versions, namely the Poincaré algebra and its non-degenerate non-Lorentzian counterparts studied in \cite{Concha:2023bly,Concha:2024dap} and listed in Appendix~\ref{sec:app}. In particular, the expanded generators are identified as
\begin{align}
    \mathtt{J}^{(0)}&\equiv J\,, & \mathtt{G}_{a}^{(0)}&\equiv G_{a}\,, & \mathtt{H}^{(0)}&\equiv H\,, & \mathtt{P}_{a}^{(0)}&\equiv P_{a}\,, \notag \\
    \mathtt{S}^{(0)}&\equiv S\,, & \mathtt{T}_{a}^{(0)}&\equiv T_{a}\,, & \mathtt{C}^{(0)}&\equiv C\,, & \mathtt{M}^{(0)}&\equiv M\,, \notag \\
    \mathtt{B}^{(0)}&\equiv B\,.
\end{align}

\begin{table}[h]
\renewcommand{\arraystretch}{1.85}
    \centering
       \begin{tabular}{|c||C{2.5cm}|C{2.5cm}|C{2.5cm}|C{2.5cm}|}
    \hline
      \rowcolor[gray]{0.9} Commutators  & $\mathfrak{B}_k$  & $\mathfrak{B}_{k}$eB  & $\mathfrak{B}_{k}$eC  & $\mathfrak{B}_{k}$eS \\ \hline
        $\left[ \mathtt{J}^{(m)},\mathtt{G}^{(n)}_{a} \right]$ & $\epsilon_{ab}\mathtt{G}^{(m+n)}_{b}$ & $\epsilon_{ab}\mathtt{G}^{(m+n)}_{b}$ & $\epsilon_{ab}\mathtt{G}^{(m+n)}_{b}$ & $\epsilon_{ab}\mathtt{G}^{(m+n)}_{b}$ \\
       \rowcolor[gray]{0.9} $\left[ \mathtt{J}^{(m)},\mathtt{P}^{(n)}_{a} \right]$ & $\epsilon_{ab}\mathtt{P}^{(m+n)}_{b}$& $\epsilon_{ab}\mathtt{P}^{(m+n)}_{b}$ & $\epsilon_{ab}\mathtt{P}^{(m+n)}_{b}$ & $\epsilon_{ab}\mathtt{P}^{(m+n)}_{b}$ \\
        $\left[ \mathtt{G}^{(m)}_{a},\mathtt{G}^{(n)}_{b} \right]$ & $-\epsilon_{ab}\mathtt{J}^{(m+n)}$ & $-\epsilon_{ab}\mathtt{S}^{(m+n)}$ &$-\epsilon_{ab}\mathtt{C}^{(m+n)}$ & $-\epsilon_{ab}\mathtt{B}^{(m+n)}$ \\
       \rowcolor[gray]{0.9} $\left[ \mathtt{H}^{(m)},\mathtt{G}^{(n)}_{a} \right]$ & $\epsilon_{ab}\mathtt{P}^{(m+n)}_{b}$ & $\epsilon_{ab}\mathtt{P}^{(m+n)}_{b}$ &$\epsilon_{ab}\mathtt{T}^{(m+n)}_{b}$ & $\epsilon_{ab}\mathtt{T}^{(m+n)}_{b}$ \\
       $\left[ \mathtt{G}^{(m)}_{a},\mathtt{P}^{(n)}_{b} \right]$ & $-\epsilon_{ab}\mathtt{H}^{(m+n)}$ & $-\epsilon_{ab}\mathtt{M}^{(m+n)}$ &$-\epsilon_{ab}\mathtt{H}^{(m+n)}$ & $-\epsilon_{ab}\mathtt{M}^{(m+n)}$ \\
       \rowcolor[gray]{0.9} $\left[ \mathtt{H}^{(m)},\mathtt{P}^{(n)}_{a} \right]$ & $\epsilon_{ab}\mathtt{G}^{(m+n+1)}_{b}$ & $\epsilon_{ab}\mathtt{G}^{(m+n+1)}_{b}$ &$\epsilon_{ab}\mathtt{G}^{(m+n+1)}_{b}$ & $\epsilon_{ab}\mathtt{G}^{(m+n+1)}_{b}$ \\
        $\left[ \mathtt{P}^{(m)}_{a},\mathtt{P}^{(n)}_{b} \right]$ & $-\epsilon_{ab}\mathtt{J}^{(m+n+1)}$ & $-\epsilon_{ab}\mathtt{S}^{(m+n+1)}$ &$-\epsilon_{ab}\mathtt{J}^{(m+n+1)}$ & $-\epsilon_{ab}\mathtt{S}^{(m+n+1)}$ \\
        \rowcolor[gray]{0.9} $\left[ \mathtt{J}^{(m)},\mathtt{T}^{(n)}_{a} \right]$  & - & - & $\epsilon_{ab}\mathtt{T}^{(m+n)}_{b}$ & $\epsilon_{ab}\mathtt{T}^{(m+n)}_{b}$\\
        $\left[ \mathtt{P}^{(m)}_{a},\mathtt{T}^{(n)}_{b} \right]$ & - & - &$-\epsilon_{ab}\mathtt{C}^{(m+n+1)}$ & $-\epsilon_{ab}\mathtt{B}^{(m+n+1)}$ \\
         \rowcolor[gray]{0.9} $\left[ \mathtt{C}^{(m)},\mathtt{P}^{(n)}_{a} \right]$ & - & - &$\epsilon_{ab}\mathtt{T}^{(m+n)}_{b}$ & $\epsilon_{ab}\mathtt{T}^{(m+n)}_{b}$ \\
     \hline
         \end{tabular}
         \captionsetup{font=footnotesize}
    \caption{Commutation relations of the $\mathfrak{B}_k$ generalization of the extended kinematical algebras.}
    \label{Table10}
\end{table}
On the other hand, the case $N=2$ reproduces the $\mathfrak{B}_{4}$ generalization of the kinematical algebra, which corresponds precisely to the Maxwellian extended kinematical algebras discussed in the previous section.

New families of kinematical symmetry algebras arise for $N\geq 3$. These algebras are neither post-Newtonian nor post-Carrollian, and they cannot be obtained as extensions of the Galilei or Carroll algebras. The particular case $N=3$ reproduces the $\mathfrak{B}_{5}$ extended kinematical algebras, whose commutation relations are listed in Appendix~\ref{sec:app2}. At the relativistic level, the $\mathfrak{B}_{5}$ algebra has proven useful to recover standard General Relativity from Chern-Simons and Born-Infeld gravity actions \cite{Edelstein:2006se,Izaurieta:2009hz,Concha:2013uhq,Concha:2014zsa}. Its non-relativistic non-degenerate counterpart is given by the $\mathfrak{B}_{5}$eB algebra (see Table~\ref{TableB1}), introduced in \cite{Concha:2020sjt} as a generalized Maxwellian exotic Bargmann algebra. In that work, its derivation was obtained through a contraction of the algebra $\mathfrak{B}_{5}\oplus\mathfrak{u}(1)^4$. The Carrollian and static versions, which had not been reported so far, are represented by the algebras $\mathfrak{B}_{5}$eC and $\mathfrak{B}_{5}$eS, respectively, as displayed in Table~\ref{TableB1}. From the commutation relations listed in Appendix~\ref{sec:app2}, it is clear that the $\mathfrak{B}_{5}$ extended kinematical algebras arise as extensions of the Maxwellian extended kinematical algebras, which correspond to the first two rows of Table~\ref{TableB1}. In particular, the first row reproduces the extended kinematical algebras introduced in \cite{Concha:2023bly}.

The $\mathfrak{B}_k$ extended kinematical algebras admit the non-vanishing components of the invariant tensor listed in Table~\ref{Table11}. The corresponding coupling constants are related to those of the starting extended kinematical algebras through the semigroup elements according to
\begin{align}
    \alpha_{i+j}&=\lambda_{i}\lambda_{j}\mu_{r}\,, & \beta_{i+j}&=\lambda_{i}\lambda_{j}\nu_{r}\,, &
    \gamma_{i+j}&=\lambda_{i}\lambda_{j}\sigma_{s}\,, &
    \zeta_{i+j}&=\lambda_{i}\lambda_{j}\rho_{s}
\end{align}
with $r=0,1$ and $s=0,1,2$. Here, the product $\lambda_{i}\lambda_{j}$ defined through the multiplication law \eqref{MLSEN} of the semigroup $S_{E}^{(N)}$.

\begin{table}[h]
\renewcommand{\arraystretch}{1.55}
    \centering
       \begin{tabular}{|c||C{2.5cm}|C{2.5cm}|C{2.5cm}|C{2.5cm}|}
    \hline
      \rowcolor[gray]{0.9} Invariant tensor  & $\mathfrak{B}_k$  & $\mathfrak{B}_{k}$eB  & $\mathfrak{B}_{k}$eC  & $\mathfrak{B}_{k}$eS \\ \hline
        $\langle \mathtt{J}^{(m)},\mathtt{J}^{(n)} \rangle$ & $-\alpha_{2n+2m}$ & 0 & $-\gamma_{2n+2m}$ & 0 \\
       \rowcolor[gray]{0.9} $\langle \mathtt{J}^{(m)},\mathtt{H}^{(n)} \rangle$ & $-\alpha_{2n+2m+1}$& 0 & $-\gamma_{2n+2m+1}$ & 0 \\
        $\langle \mathtt{H}^{(m)},\mathtt{H}^{(n)} \rangle$& $-\alpha_{2n+2m+2}$ & 0 & $-\gamma_{2n+2m+2}$ & 0 \\
       \rowcolor[gray]{0.9} $\langle \mathtt{G}_{a}^{(m)},\mathtt{G}_{b}^{(n)} \rangle$& $\alpha_{2n+2m}\,\delta_{ab}$ & $\beta_{2n+2m}\,\delta_{ab}$ & $\gamma_{2n+2m}\,\delta_{ab}$ & $\zeta_{2n+2m}\,\delta_{ab}$ \\
       $\langle \mathtt{G}_{a}^{(m)},\mathtt{P}_{b}^{(n)} \rangle$& $\alpha_{2n+2m+1}\,\delta_{ab}$ & $\beta_{2n+2m+1}\,\delta_{ab}$ & $\gamma_{2n+2m+1}\,\delta_{ab}$ & $\zeta_{2n+2m+1}\,\delta_{ab}$ \\
       \rowcolor[gray]{0.9} $\langle \mathtt{P}_{a}^{(m)},\mathtt{P}_{b}^{(n)} \rangle$& $\alpha_{2n+2m+2}\,\delta_{ab}$ & $\beta_{2n+2m+2}\,\delta_{ab}$ & $\gamma_{2n+2m+2}\,\delta_{ab}$ & $\zeta_{2n+2m+2}\,\delta_{ab}$ \\
       $\langle \mathtt{J}^{(m)},\mathtt{S}^{(n)} \rangle$ & - & $-\beta_{2n+2m}$ & - & $-\zeta_{2n+2m}$ \\
       \rowcolor[gray]{0.9} $\langle \mathtt{J}^{(m)},\mathtt{M}^{(n)} \rangle$ & - & $-\beta_{2n+2m+1}$ & - & $-\zeta_{2n+2m+1}$ \\
        $\langle \mathtt{H}^{(m)},\mathtt{M}^{(n)} \rangle$ & - & $-\beta_{2n+2m+2}$ & - & $-\zeta_{2n+2m+2}$ \\
        \rowcolor[gray]{0.9} $\langle \mathtt{H}^{(m)},\mathtt{S}^{(n)} \rangle$ & - & $-\beta_{2n+2m+1}$ & - & $-\zeta_{2n+2m+1}$ \\
        $\langle \mathtt{J}^{(m)},\mathtt{C}^{(n)} \rangle$ & - & - & $-\gamma_{2n+2m}$ & 0 \\
         \rowcolor[gray]{0.9} $\langle \mathtt{J}^{(m)},\mathtt{B}^{(n)} \rangle$ & - & - & - & $-\zeta_{2n+2m}$ \\
          $\langle \mathtt{P}_{a}^{(m)},\mathtt{T}_{b}^{(n)} \rangle$& - & - & $\gamma_{2n+2m+2}\,\delta_{ab}$ & $\zeta_{2n+2m+
          2}\,\delta_{ab}$ \\
     \hline
         \end{tabular}
         \captionsetup{font=footnotesize}
    \caption{Non-vanishing components of the invariant tensor for the $\mathfrak{B}_k$ extended kinematical algebras.}
    \label{Table11}
\end{table}
One can observe that the components proportional to $\alpha_i$, with $i=0,\ldots,N-1$, reproduce the invariant tensor structure of the $\mathfrak{B}_{N+1}$ algebra. Since $N=k-2$, this recursive property allows us to express the CS gravity action for the $\mathfrak{B}_{k}$ algebra as the CS action for $\mathfrak{B}_{k-1}$ plus an additional contribution proportional to $\alpha_{N}$:
\begin{align}
    I_{\mathfrak{B}_{k}}
    =\frac{k}{4\pi}\int\left(\mathcal{L}_{\mathfrak{B}_{k-1}}+\alpha_{N}\mathcal{L}_{N}\right)\,.
\end{align}
An analogous recursive behavior holds for the non-Lorentzian counterparts of the $\mathfrak{B}_{k}$ algebra.


\section{Discussion}\label{sec5}

In this work, we have presented a Maxwellian extension of the kinematical Lie algebras by promoting the contraction process of the original Bacry and Lévy-Leblond cube \cite{Bacry:1968zf} to an S-expansion mechanism, with $S_{E}^{(2)}$ as the relevant semigroup. We have shown that the non-degenerate non-Lorentzian Maxwell gravity theories previously constructed in the literature arise naturally within this unified framework. Moreover, we have shown that both the original cube \cite{Bacry:1968zf} and the Maxwellian counterpart belong to an infinite hierarchy of generalized kinematical algebras obtained through arbitrary $S_{E}^{(N)}$ semigroups, where the case $N=1$ reproduces the Bacry and Lévy-Leblond cube. These generalized structures correspond to the so-called $\mathfrak{B}_{k}$ algebras and their non-Lorentzian regimes. Remarkably, the present approach also provides the non-vanishing components of the invariant tensor required for the construction of CS gravity actions for both Lorentzian and non-Lorentzian symmetry algebras.

The results presented here open several avenues for further investigation. At the gravitational level, it would be interesting to study the dynamics associated with the Maxwell kinematical algebras. In particular, it remains to clarify the physical role of the additional gauge fields appearing in the non-Lorentzian regimes of the Maxwell cube. These fields may admit an interpretation in terms of post-Newtonian or post-Carrollian corrections, in analogy with higher-order expansions of relativistic gravity. It is also natural to explore whether the additional non-Lorentzian gauge fields can be related to gravito-magnetic-like effects \cite{Jantzen:1992rg}. In this context, one may ask whether they contribute to generalized notions of torsion, non-inertial forces, or effective background fluxes in non-relativistic and Carrollian geometries. From this perspective, the Maxwellian non-Lorentzian algebras introduced here may provide a novel arena to investigate generalized gravitational interactions beyond the standard kinematical frameworks.

In the generalized kinematical setting, it would be worth analyzing the space of classical solutions of the corresponding $\mathfrak{B}_{k}$ CS gravities, including black-hole configuration, cosmological backgrounds, and their thermodynamic properties. The additional content is expected to modify both the global structure of solutions and the associated conserved charges, potentially leading to new thermodynamic contributions. The first non-trivial case beyond the Maxwellian level, corresponding to the $\mathfrak{B}_5$ algebra, deserves special attention. At the relativistic level, $\mathfrak{B}_5$ has been shown to play a central role in recovering General Relativity in suitable limits \cite{Edelstein:2006se,Izaurieta:2009hz,Concha:2013uhq,Concha:2014zsa}. It would be highly interesting to investigate whether its non-Lorentzian counterparts encode analogous subleading gravitational structures and to elucidate the geometric and dynamical interpretation of the corresponding gauge fields.

On a more conceptual level, it would be interesting to explore whether the non-Lorentzian algebras presented here admit realization as asymptotic symmetries of gravitational models, and whether they give rise to new extensions of $\mathfrak{bms}_{3}$ algebra. On the other hand, motivated by recent developments in three-dimensional AdS-Carroll gravity, where the asymptotic symmetry algebra corresponds to an infinite-dimensional extension of a generalized Maxwell algebra \cite{Aviles:2025ygw}, one could explore whether infinite-dimensional extensions of the $\mathfrak{B}_{k}$ algebras emerge in a symilar way. This could provide new insights into the role of non-Lorentzian limits within holography.

Finally, another natural direction concerns the extension of our results to supersymmetric and higher-spin theories. In particular, applying the $S$-expansion method to non-Lorentzian regimes could lead to new classes of non-degenerate supergravity theories. While supersymmetric extensions of the Bacry and Lévy-Leblond cube have been recently constructed in \cite{Concha:2024dap}, it would be interesting to generalize the present approach by considering different starting (super)algebras beyond the (super) AdS case. At the higher-spin level, it would be worth exploring whether spin-$5/2$ symmetry algebras \cite{Aragone:1983sz,Fuentealba:2015jma,Henneaux:2015ywa,Caroca:2023oie} can be incorporated into generalized kinematical frameworks along the lines of \cite{Bergshoeff:2016soe}.

\section*{Acknowledgments}

This work was funded by the National Agency for Research and Development ANID - FONDECYT grants No. 1250642 and 1231133. This work was supported by DIREG 04/2025 of the Universidad Católica de la Santísima Concepción (P.C.). P.C. and E.R. would like to thank the Dirección de Investigación and Vice-rectoría de Investigación of the Universidad Católica de la Santísima Concepción, Chile, for their constant support.

\appendix

\section{Explicit commutation relations of extended kinematical algebras}\label{sec:app}
This appendix contains the tables with the complete list of commutation relations of the extended kinematical algebras appearing in \cite{Concha:2023bly}.
\begin{table}[h]
    \centering
    \begin{tabular}{|l||r|r|r|r|}
    \hline
      \rowcolor[gray]{0.8} Commutators  & AdS  & Poincaré  & Extended  & Extended \\ 
      \rowcolor[gray]{0.8} &  &  & Newton-Hooke & Bargmann \\ \hline
        $\left[ J,G_{a} \right]$ & $\epsilon_{ab}G_{b}$ & $\epsilon_{ab}G_{b}$& $\epsilon_{ab}G_{b}$& $\epsilon_{ab}G_{b}$\\
        \rowcolor[gray]{0.8}$\left[ J,P_{a} \right]$ & $\epsilon_{ab}P_{b}$& $\epsilon_{ab}P_{b}$ & $\epsilon_{ab}P_{b}$ & $\epsilon_{ab}P_{b}$ \\
        $\left[ G_{a},G_{b} \right]$ & $-\epsilon_{ab}J$ & $-\epsilon_{ab}J$ &$-\epsilon_{ab}S$ & $-\epsilon_{ab}S$ \\
        \rowcolor[gray]{0.8}$\left[ H,G_{a} \right]$ & $\epsilon_{ab}P_{b}$ & $\epsilon_{ab}P_{b}$ &$\epsilon_{ab}P_{b}$ & $\epsilon_{ab}P_{b}$ \\
        $\left[ G_{a},P_{b} \right]$ & $-\epsilon_{ab}H$ & $-\epsilon_{ab}H$ &$-\epsilon_{ab}M$ & $-\epsilon_{ab}M$ \\
        \rowcolor[gray]{0.8}$\left[ H,P_{a} \right]$ & $\epsilon_{ab}G_{b}$ & $0$ &$\epsilon_{ab}G_{b}$ & $0$ \\
        $\left[ P_{a},P_{b} \right]$ & $-\epsilon_{ab}J$ & $0$ &$-\epsilon_{ab}S$ & $0$ \\
     \hline
         \end{tabular}
         \captionsetup{font=footnotesize}
    \caption{Commutation relations of the AdS, Poincaré, extended Newton-Hooke and extended Bargmann algebras.}
    \label{TableA1}
\end{table}

\begin{table}[h]
    \centering
    \begin{tabular}{|l||r|r|r|r|}
    \hline
      \rowcolor[gray]{0.8} Commutators  & Extended  & Extended  & Extended  & Extended \\ 
      \rowcolor[gray]{0.8} & AdS-Carroll  & Carroll & AdS-Static & Static \\ \hline
        $\left[ J,G_{a} \right]$ & $\epsilon_{ab}G_{b}$ & $\epsilon_{ab}G_{b}$& $\epsilon_{ab}G_{b}$& $\epsilon_{ab}G_{b}$\\
        \rowcolor[gray]{0.8}$\left[ J,P_{a} \right]$ & $\epsilon_{ab}P_{b}$& $\epsilon_{ab}P_{b}$ & $\epsilon_{ab}P_{b}$ & $\epsilon_{ab}P_{b}$ \\
        $\left[ G_{a},G_{b} \right]$ & $-\epsilon_{ab}C$ & $-\epsilon_{ab}C$ &$-\epsilon_{ab}B$ & $-\epsilon_{ab}B$ \\
        \rowcolor[gray]{0.8}$\left[ H,G_{a} \right]$ & $\epsilon_{ab}T_{b}$ & $\epsilon_{ab}T_{b}$ &$\epsilon_{ab}T_{b}$ & $\epsilon_{ab}T_{b}$ \\
        $\left[ G_{a},P_{b} \right]$ & $-\epsilon_{ab}H$ & $-\epsilon_{ab}H$ &$-\epsilon_{ab}M$ & $-\epsilon_{ab}M$ \\
        \rowcolor[gray]{0.8}$\left[ H,P_{a} \right]$ & $\epsilon_{ab}G_{b}$ & $0$ &$\epsilon_{ab}G_{b}$ & $0$ \\
        $\left[ P_{a},P_{b} \right]$ & $-\epsilon_{ab}J$ & $0$ &$-\epsilon_{ab}S$ & $0$ \\
         \rowcolor[gray]{0.8}$\left[ J,T_{a} \right]$ & $\epsilon_{ab}T_{b}$ & $\epsilon_{ab}T_{b}$ &$\epsilon_{ab}T_{b}$ & $\epsilon_{ab}T_{b}$ \\
           $\left[ P_{a},T_{b} \right]$ & $-\epsilon_{ab}C$ & $0$ &$-\epsilon_{ab}B$ & $0$ \\
           \rowcolor[gray]{0.8}$\left[ C,P_{a} \right]$ & $\epsilon_{ab}T_{b}$ & $\epsilon_{ab}T_{b}$ &$\epsilon_{ab}T_{b}$ & $\epsilon_{ab}T_{b}$ \\
     \hline
         \end{tabular}
         \captionsetup{font=footnotesize}
    \caption{Commutation relations for the extended versions of the AdS–Carroll, Carroll, AdS–Static, and Static algebras.}
    \label{TableA2}
\end{table}
\section{Explicit commutation relations of the $\mathfrak{B}_{5}$ extended kinematical algebras}\label{sec:app2}
This appendix contains the table with the complete list of commutators of the $\mathfrak{B}_{5}$ algebra and its non-Lorentzian counterpart. 
\begin{table}[h]
    \centering
       \begin{tabular}{|c||C{2.5cm}|C{2.5cm}|C{2.5cm}|C{2.5cm}|}
    \hline
      \rowcolor[gray]{0.9} Commutators  & $\mathfrak{B}_5$  & $\mathfrak{B}_{5}$eB  & $\mathfrak{B}_{5}$eC  & $\mathfrak{B}_{5}$eS \\ \hline
        $\left[ \mathtt{J},\mathtt{G}_{a} \right]$ & $\epsilon_{ab}\mathtt{G}_{b}$ & $\epsilon_{ab}\mathtt{G}_{b}$ & $\epsilon_{ab}\mathtt{G}_{b}$ & $\epsilon_{ab}\mathtt{G}_{b}$ \\
        $\left[ \mathtt{J},\mathtt{P}_{a} \right]$ & $\epsilon_{ab}\mathtt{P}_{b}$& $\epsilon_{ab}\mathtt{P}_{b}$ & $\epsilon_{ab}\mathtt{P}_{b}$ & $\epsilon_{ab}\mathtt{P}_{b}$ \\
        $\left[ \mathtt{G}_{a},\mathtt{G}_{b} \right]$ & $-\epsilon_{ab}\mathtt{J}$ & $-\epsilon_{ab}\mathtt{S}$ &$-\epsilon_{ab}\mathtt{C}$ & $-\epsilon_{ab}\mathtt{B}$ \\
        $\left[ \mathtt{H},\mathtt{G}_{a} \right]$ & $\epsilon_{ab}\mathtt{P}_{b}$ & $\epsilon_{ab}\mathtt{P}_{b}$ &$\epsilon_{ab}\mathtt{T}_{b}$ & $\epsilon_{ab}\mathtt{T}_{b}$ \\
       $\left[ \mathtt{G}_{a},\mathtt{P}_{b} \right]$ & $-\epsilon_{ab}\mathtt{H}$ & $-\epsilon_{ab}\mathtt{M}$ &$-\epsilon_{ab}\mathtt{H}$ & $-\epsilon_{ab}\mathtt{M}$ \\
       $\left[ \mathtt{J},\mathtt{T}_{a} \right]$  & - & - & $\epsilon_{ab}\mathtt{T}_{b}$ & $\epsilon_{ab}\mathtt{T}_{b}$\\
         $\left[ \mathtt{C},\mathtt{P}_{a} \right]$ & - & - &$\epsilon_{ab}\mathtt{T}_{b}$ & $\epsilon_{ab}\mathtt{T}_{b}$ \\ \hline
       \rowcolor[gray]{0.9} $\left[ \mathtt{H},\mathtt{P}_{a} \right]$ & $\epsilon_{ab}\mathtt{Z}_{b}$ & $\epsilon_{ab}\mathtt{Z}_{b}$ &$\epsilon_{ab}\mathtt{Z}_{b}$ & $\epsilon_{ab}\mathtt{Z}_{b}$ \\
       \rowcolor[gray]{0.9} $\left[ \mathtt{P}_{a},\mathtt{P}_{b} \right]$ & $-\epsilon_{ab}\mathtt{Z}$ & $-\epsilon_{ab}\mathtt{T}$ &$-\epsilon_{ab}\mathtt{Z}$ & $-\epsilon_{ab}\mathtt{T}$ \\
        \rowcolor[gray]{0.9} $\left[ \mathtt{J},\mathtt{Z}_{a} \right]$ & $\epsilon_{ab}\mathtt{Z}_{b}$ & $\epsilon_{ab}\mathtt{Z}_{b}$ & $\epsilon_{ab}\mathtt{Z}_{b}$ & $\epsilon_{ab}\mathtt{Z}_{b}$ \\
        \rowcolor[gray]{0.9} $\left[ \mathtt{Z},\mathtt{G}_{a} \right]$ & $\epsilon_{ab}\mathtt{Z}_{b}$ & $\epsilon_{ab}\mathtt{Z}_{b}$ & $\epsilon_{ab}\mathtt{Z}_{b}$ & $\epsilon_{ab}\mathtt{Z}_{b}$ \\
        \rowcolor[gray]{0.9} $\left[ \mathtt{G}_{a},\mathtt{Z}_{b} \right]$ & $-\epsilon_{ab}\mathtt{Z}$ & $-\epsilon_{ab}\mathtt{T}$ &$-\epsilon_{ab}\mathtt{L}$ & $-\epsilon_{ab}\mathtt{Y}$ \\
        \rowcolor[gray]{0.9} $\left[ \mathtt{P}_{a},\mathtt{T}_{b} \right]$ & - & - &$-\epsilon_{ab}\mathtt{L}$ & $-\epsilon_{ab}\mathtt{Y}$ \\ \hline
        \rowcolor[gray]{0.75}$\left[ \mathtt{J},\mathtt{N}_{a} \right]$ & $\epsilon_{ab}\mathtt{N}_{b}$ & $\epsilon_{ab}\mathtt{N}_{b}$ & $\epsilon_{ab}\mathtt{N}_{b}$ & $\epsilon_{ab}\mathtt{N}_{b}$ \\
        \rowcolor[gray]{0.75}$\left[ \mathtt{Z},\mathtt{P}_{a} \right]$ & $\epsilon_{ab}\mathtt{N}_{b}$ & $\epsilon_{ab}\mathtt{N}_{b}$ & $\epsilon_{ab}\mathtt{N}_{b}$ & $\epsilon_{ab}\mathtt{N}_{b}$ \\
        \rowcolor[gray]{0.75} $\left[ \mathtt{H},\mathtt{Z}_{a} \right]$ & $\epsilon_{ab}\mathtt{N}_{b}$ & $\epsilon_{ab}\mathtt{N}_{b}$ & $\epsilon_{ab}\mathtt{L}_{b}$ & $\epsilon_{ab}\mathtt{L}_{b}$ \\
        \rowcolor[gray]{0.75} $\left[ \mathtt{Z}_{a},\mathtt{P}_{b} \right]$ & $-\epsilon_{ab}\mathtt{N}$ & $-\epsilon_{ab}\mathtt{V}$ &$-\epsilon_{ab}\mathtt{N}$ & $-\epsilon_{ab}\mathtt{V}$ \\
        \rowcolor[gray]{0.75} $\left[ \mathtt{G}_{a},\mathtt{N}_{b} \right]$ & $-\epsilon_{ab}\mathtt{N}$ & $-\epsilon_{ab}\mathtt{V}$ &$-\epsilon_{ab}\mathtt{N}$ & $-\epsilon_{ab}\mathtt{V}$ \\
        \rowcolor[gray]{0.75}   $\left[ \mathtt{J},\mathtt{L}_{a} \right]$  & - & - & $\epsilon_{ab}\mathtt{L}_{b}$ & $\epsilon_{ab}\mathtt{L}_{b}$\\
        \rowcolor[gray]{0.75}   $\left[ \mathtt{Z},\mathtt{T}_{a} \right]$  & - & - & $\epsilon_{ab}\mathtt{L}_{b}$ & $\epsilon_{ab}\mathtt{L}_{b}$\\
        \rowcolor[gray]{0.75} $\left[ \mathtt{C},\mathtt{N}_{a} \right]$ & - & - &$\epsilon_{ab}\mathtt{L}_{b}$ & $\epsilon_{ab}\mathtt{L}_{b}$ \\ 
        \rowcolor[gray]{0.75} $\left[ \mathtt{L},\mathtt{P}_{a} \right]$ & - & - &$\epsilon_{ab}\mathtt{L}_{b}$ & $\epsilon_{ab}\mathtt{L}_{b}$ \\ 
     \hline
         \end{tabular}
         \captionsetup{font=footnotesize}
    \caption{Commutation relations of the $\mathfrak{B}_5$ extended kinematical algebras.}
    \label{TableB1}
\end{table}
Here, the generators have been identified in terms of the expanded generators listed in Table~\ref{Table9} as
\begin{align}
    \mathtt{J}^{(0)}&\equiv\mathtt{J}\,, & \mathtt{J}^{(1)}&\equiv \mathtt{Z}\,, & \mathtt{G}^{(0)}_{a}&\equiv \mathtt{G}_{a}\,, & \mathtt{G}^{(1)}_{a}&\equiv \mathtt{Z}_{a}\,, \notag\\
    \mathtt{H}^{(0)}&\equiv\mathtt{H}\,, & \mathtt{H}^{(1)}&\equiv \mathtt{N}\,, & \mathtt{P}^{(0)}_{a}&\equiv \mathtt{P}_{a}\,, & \mathtt{P}^{(1)}_{a}&\equiv \mathtt{N}_{a}\,, \notag\\
    \mathtt{S}^{(0)}&\equiv\mathtt{S}\,, & \mathtt{S}^{(1)}&\equiv \mathtt{T}\,, & \mathtt{T}^{(0)}_{a}&\equiv \mathtt{T}_{a}\,, & \mathtt{T}^{(1)}_{a}&\equiv \mathtt{L}_{a}\,, \notag\\
    \mathtt{C}^{(0)}&\equiv\mathtt{C}\,, & \mathtt{C}^{(1)}&\equiv \mathtt{L}\,, & \mathtt{B}^{(0)}&\equiv \mathtt{B}\,, & \mathtt{B}^{(1)}&\equiv \mathtt{Y}\,, \notag\\
    \mathtt{M}^{(0)}&\equiv\mathtt{M}\,, & \mathtt{M}^{(1)}&\equiv \mathtt{V}\,.
\end{align}

\bibliographystyle{fullsort.bst}
 
\bibliography{3D_gravity_based_on_Maxwellian_kinematical_algebras}

\end{document}